\documentclass[aps,physrev,preprint,superscriptaddress]{revtex4-2}
\usepackage{graphicx}
\usepackage{booktabs}   
\usepackage{multirow}   
\usepackage{caption}    
\usepackage{ragged2e}
\usepackage{amsmath}
\usepackage{enumitem}
\usepackage{xcolor}
\usepackage{hyperref}
\usepackage{xcolor}

\hypersetup{
    colorlinks=true,
    linkcolor=blue,      
    citecolor=blue,
    urlcolor=blue
}
\usepackage[english]{babel}
\begin{document}
\renewcommand{\figurename}{FIG.}

\title{Entropy-Driven Sensor Deployment and Source Detection in Hypergraphs}


\author{Qiao Ke}
\affiliation{Alibaba Research Center for Complexity Sciences, Hangzhou Normal University, Hangzhou 311121, China}

\author{Chengjun Zhang}
\affiliation{Alibaba Research Center for Complexity Sciences, Hangzhou Normal University, Hangzhou 311121, China}

\author{Chuang Liu}
\affiliation{Alibaba Research Center for Complexity Sciences, Hangzhou Normal University, Hangzhou 311121, China}

\author{Mingxia Jing}
\email{Mingxiajing163@163.com}
\affiliation{Key Laboratory for Prevention and Control of Emerging Infectious Diseases and Public Health Security, the Xinjiang Production and Construction Corps, Shihezi University School of Public Health, Shihezi, Xinjiang 832000, China}

\author{Suoyi Tan}
\email{tansuoyi@nudt.edu.cn}
\affiliation{College of Systems Engineering, National University of Defense Technology, Changsha 410073, China}

\author{Xiu-Xiu Zhan}
\email{zhanxiuxiu@hznu.edu.cn}
\affiliation{Alibaba Research Center for Complexity Sciences, Hangzhou Normal University, Hangzhou 311121, China}
\affiliation{Engineering Research Center of Mobile Health Management System, Ministry of Education, Hangzhou 311121, China}

\date{\today}

\begin{abstract}
Identifying the diffusion source in complex networks is critical for understanding and controlling epidemic spread. In realistic settings, full observation of node states is rarely available, making sensor-based source detection a practical alternative. However, existing sensor-based methods are often confined to simple networks, failing to capture the higher-order group dynamics of real-world spreading process. By deploying a limited number of sensors to monitor the diffusion process, one can infer the origin from partial observations. Yet, determining optimal sensor placement is challenging, i.e., poor deployment leads to redundant or noisy data, while optimal placement must balance coverage diversity and information value under limited resources. To address these challenges, we propose a dedicated framework termed Sensor-based Source Detection in Hypergraphs (SSDH). Specifically, we introduce a novel entropy-driven sensor deployment strategy that effectively captures critical early-stage diffusion signals by maximizing information gain under limited resources. Furthermore, we develop a source localization algorithm that quantifies propagation uncertainty through a newly defined path uncertainty-based score. By integrating this score with topological distance, SSDH enables accurate and robust source identification.
Extensive experiments on both synthetic and empirical hypergraphs demonstrate that SSDH consistently outperforms competing algorithms by 5\%–30\% across different sensor ratios, final spreading ratios, and infection probabilities. These results validate the effectiveness of SSDH and highlight its superior capability to tackle source localization in complex systems characterized by higher-order interactions.
\end{abstract}


\maketitle

\section{Introduction}

Identifying the infection source, often referred to as patient zero, is fundamental for understanding and controlling epidemic outbreaks \cite{williams2005impact,smith2006responding,zhan2025measuring,fanelli2020analysis,mckay2017patient,yao2023lesson}. The source detection problem aims to infer the origin of contagion from observed infection information across a networked population \cite{antulov2015identification,ru2023inferring,zhang2025locating}. Depending on the availability and completeness of observations \cite{jiang2016identifying}, existing studies generally classify this problem into three categories: (i) complete observations, where both the network topology and node states are fully known; (ii) partial observations, where certain components of the system are unobserved, commonly appearing as incomplete knowledge of the network topology or missing infection states of some nodes; and (iii) sensor-based observations, where only data from deployed sensors are accessible. Among these settings \cite{lokhov2014inferring,chang2018maximum,li2019locating,luo2013estimating,zhu2014information}, sensor-based approaches have attracted growing attention due to their practicality and scalability in real-world monitoring systems \cite{paluch2020optimizing}. Given that our observations are limited to sensor data, it becomes essential to maximize the utility of the information available under partial observability \cite{li2011attack}. Such methods typically involve two key components, i.e., the optimal deployment of sensors and the accurate identification of infection sources based on sensor feedback.

In recent years, extensive research has been devoted to sensor deployment and source detection in simple networks, where interactions occur only between node pairs. Effective sensor placement serves as the cornerstone of accurate source inference, with strategies evolving from random \cite{pinto2012locating} and topology-driven approaches \cite{ali2020revisit} to advanced optimization-based methods that minimize propagation distance \cite{wang2022rapid}, maximize information entropy \cite{hu2023source}, or enhance network coverage \cite{cheng2024sdsi,cheng2025efficient}.
Parallel to this, source detection algorithms have undergone substantial development since the seminal Pinto-Thiran–Vetterli (PTV) model, which first leveraged sensor timestamps for localization \cite{pinto2012locating}. Subsequent studies have relaxed the single-path assumption of PTV by incorporating multi-path propagation \cite{paluch2020enhancing}, alternative path construction schemes \cite{tang2018estimating}, or novel paradigms such as correlation-based inference \cite{wang2020locating} and path-recording mechanisms \cite{zhu2022locating}. Other research efforts have emphasized improving computational efficiency \cite{paluch2018fast}, exploiting directional diffusion information \cite{yang2020locating}, enabling early-stage outbreak localization \cite{wang2023lightweight,wang2022rapid}, and addressing detection under incomplete observation scenarios \cite{cheng2024sdsi}.

Despite these advances, most existing source detection methods remain confined to simple network representations and thus fail to capture the inherently higher-order, group-based dynamics observed in real-world systems such as disease transmission within households, classrooms, or public transportation networks \cite{pan2024robustness,kim2024higher,bick2023higher,majhi2022dynamics}. Accurately modeling such collective diffusion processes requires a framework capable of representing multi-node interactions, for which hypergraphs offer a natural and mathematically rigorous formulation.
In light of this insight, several recent studies have extended classical models from simple networks to hypergraphs \cite{jhun2021effective,feng2025hypergraph}. For instance, Yu et al. \cite{yu2024source} and Ke et al. \cite{ke2025source} generalized the Dynamic Message Passing (DMP) framework to hypergraph structures, enabling the tracing of epidemic spread under higher-order interactions. Cheng et al. \cite{cheng2025hyperdet} further introduced a hypergraph neural network-based framework for source identification. However, these approaches assume complete observability of node states and network topology, leaving the problem of sensor-based source detection in hypergraphs largely unexplored. Extending sensor-based approaches from simple networks to hypergraphs presents substantial challenges. Fundamental notions such as distance and reachability must be redefined to accommodate the unique topological structure of hypergraphs, while group-based propagation dynamics introduce additional uncertainty in path inference and render conventional deployment strategies suboptimal \cite{xu2023iqabc}.
To bridge this gap, we conduct the first systematic investigation of sensor-based source detection on hypergraphs and propose the Sensor-based Source Detection in Hypergraphs (SSDH) framework, offering a new paradigm for monitoring and controlling diffusion in systems characterized by higher-order interactions. Specifically, sensor deployment in hypergraphs poses a particularly challenging problem due to the combinatorial nature of higher-order connections and the need to balance coverage diversity with information utility. To address this challenge, we design an entropy-driven sensor deployment strategy, which leverages the principle of diminishing marginal returns to achieve both efficiency and robustness. We further introduce a source detection algorithm that combines an efficient candidate filtering mechanism with a path entropy-based metric. By coupling adaptive path uncertainty with propagation distance through a tunable weighting scheme, SSDH enables accurate, interpretable, and scalable source localization across diverse hypergraphs.

\section{Preliminary}
\subsection{Problem Definition}
We begin by formally defining the sensor-based source detection problem in hypergraphs. A hypergraph is represented as $H=(V,E)$, where $V = \{ v_1, v_2, \cdots, v_N \}$ is the set of nodes, and $E = \{ e_1, e_2, \cdots, e_M \}$ is the set of hyperedges. Each hyperedge $e_i \subseteq E$ is a set that contains a certain number of nodes, i.e., $e_i = \{ v_{i1}, v_{i2}, \cdots, v_{ik} \}$. The corresponding simple network of $H$ can be denoted as $G=(V,E')$, where an edge $e'=(v_i, v_j)\in E'$ exists if and only if the two nodes $v_i$ and $v_j$ co-occur in at least one hyperedge of $H$. Accordingly, the distance between any pair of nodes in $H$ is defined as the length of the shortest path connecting them in $G$. This distance metric serves as a foundational component for the subsequent sensor deployment strategies and the design of the source identification algorithm.
A subset of nodes is selected as sensors, denoted by $O = \{ o_1, o_2, \cdots, o_m \}$, which record infection-related observations during the diffusion process. We assume that the contagion originates from a single source node $v_s \in V$. At time $t=0$, the source node $v_s$ is assumed to be infected. As the contagion spreads through the hypergraph over discrete time steps, each sensor node that becomes infected records both its infection time and the identity of the node that transmitted the infection. Once the cumulative number of infected nodes surpasses a predefined threshold $\theta$, the observation time is denoted as $T_{\theta}$. At this stage, the objective is to identify the initial infection source by leveraging the infection timestamps recorded by the sensors together with the underlying hypergraph topology.
Accordingly, the sensor-based source localization problem in a hypergraph can be formulated as identifying the source node $v_{\hat{s}}$ that maximizes the following likelihood function:
\begin{equation}
    v_{\hat{s}} = \arg\max_{v_s \in V} P(Q|H, v_s),
\end{equation}
where $Q$ denotes the sensor observations collected until $T_{\theta}$, including each sensor’s infection state, its recorded infection time, and the identity of the node that transmitted the infection to it, i.e., hereafter referred to as the infector of that sensor.

\subsection{Hypergraph-based SI Spreading Model}

The spreading process is modeled through a discrete-time Susceptible-Infected (SI) model~\cite{xie2023efficient} specifically adapted for hypergraphs. The process starts at $t=0$ with a single infected source node. During each subsequent time step, every infected node randomly selects one of its incident hyperedges and attempts to infect all the susceptible nodes within that hyperedge with probability $\lambda$. The process continues until the proportion of infected nodes exceeds a predefined threshold $\theta$, which serves as a tunable termination parameter.

\begin{figure}
    \centering
    \includegraphics[width=\linewidth]{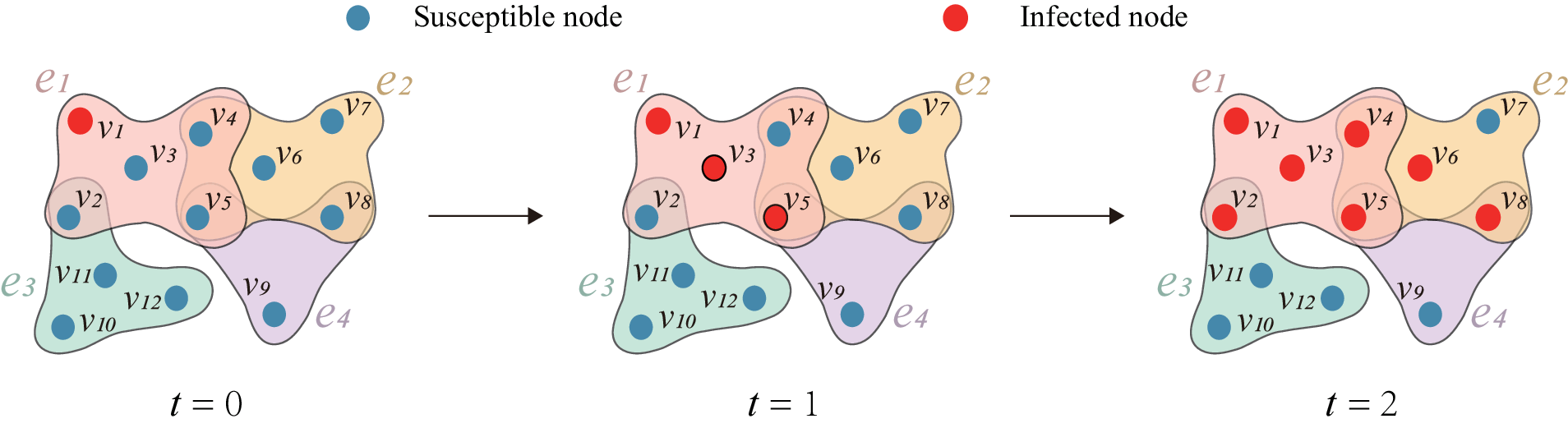}
    \caption{Example of the hypergraph-based SI spreading process. }
    \label{fig:spread}
\end{figure}

A detailed example is illustrated in Fig.~\ref{fig:spread}. At $t=0$, node $v_1$ is initially infected. During the first time step ($t=0\rightarrow 1$), $v_1$ activates hyperedge $e_1$ and transmits the infection to nodes $v_3$ and $v_5$. In the second time step ($t=1\rightarrow 2$), all infected nodes attempt to further propagate the infection: $v_1$ reactivates along $e_1$ and infects $v_2$; $v_3$ activates along the same hyperedge $e_1$ to infect $v_4$; meanwhile, $v_5$ activates along $e_2$, resulting in the infection of $v_6$ and $v_8$.

\section{Sensor-based Source Detection in Hypergraphs (SSDH)}
\begin{figure}
    \centering
    \includegraphics[width=\linewidth]{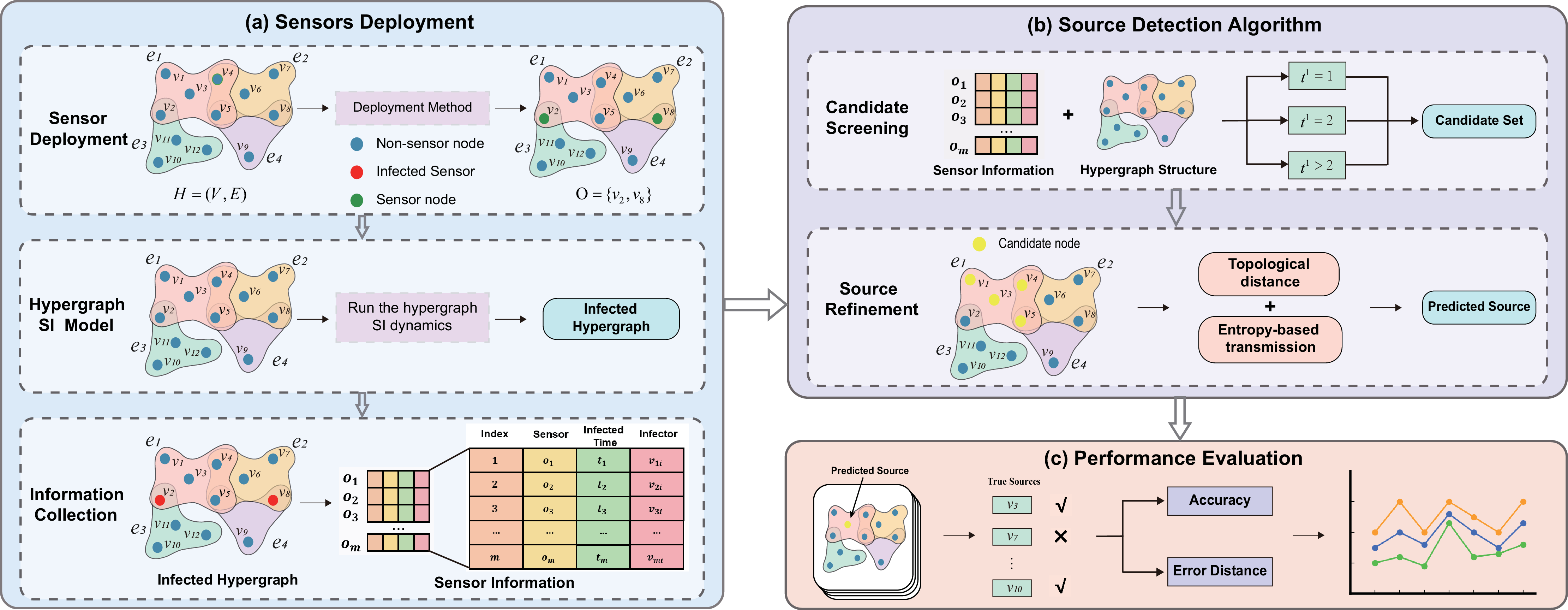}
    \caption{Overall workflow of the SSDH algorithm. (a) A proportion of nodes is selected as sensor nodes using the entropy-based deployment strategy, and their infection information is extracted from the resulting infected hypergraph. (b) Candidate screening based on the earliest sensor infection time $t^1$, followed by source refinement through the integration of topological distance and entropy-based transmission. (c) The effectiveness of the proposed algorithm is evaluated through comparative experiments.}
    \label{fig:framework}
\end{figure}

In this section,  we present the details of our sensor-based source detection framework, which comprises two key components, i.e., the sensor deployment strategy and the source detection algorithm. The overall workflow of the proposed framework is illustrated in Fig.~\ref{fig:framework}.

\subsection{Entropy-based Sensor Deployment Strategy}
We propose an entropy-based sensor deployment strategy to optimally distribute sensors in a hypergraph. The method quantifies the information gain achieved by adding a new sensor from an information-theoretic perspective, ensuring both robustness and efficient coverage. By introducing controlled redundancy in structurally critical regions, it enhances network stability while maintaining diminishing marginal returns. The core algorithm adopts an iterative greedy selection process that sequentially deploys sensors based on their contribution to uncertainty reduction. Specifically, we define the marginal gain $\Delta C(e)$ of adding a new sensor associated with hyperedge $e$, which currently contains $m_e$ sensors, as the change in its entropy-driven logarithmic utility:

\begin{equation}
    \Delta C(e) = \log(m_e+2) - \log(m_e+1).
\end{equation}
The main procedures of the proposed deployment strategy are outlined as follows:

\begin{enumerate}[label=(\roman*)]
    \item Initially, the sensor set $O$ is empty, and the number of sensors $m_e$ in each hyperedge $e$ is initialized to $0$.
    
    \item For any candidate sensor node $v_i \in V\setminus O$ added in the hypergraph, its uncertainty reduction $\Phi(i)$ is defined as the total reduction contributed across all incident hyperedges, formulated as
    \begin{equation}
        \Phi(i) = \sum_{e \in E_i} \Delta C(e),
    \end{equation}
    where $E_i$ denotes the set of hyperedges containing node $v_i$. The node $v_{i^*}$ that yields the maximum value of uncertainty reduction, i.e.,  $\Phi(i^*)$, is then selected and added to the sensor set $O$. Subsequently, the sensor counts (i.e., the value of $m_e$) of all hyperedges connected to $v_{i^*}$ are updated to reflect the new deployment state.

    \item The iterative process described in Step~(ii) is repeated until the sensor ratio, defined as the number of selected sensors divided by the total number of nodes in the hypergraph ($m/N$), reaches the predefined budget constraint. 
\end{enumerate}

\begin{figure}[htp!]
    \centering
    \includegraphics[width=\linewidth]{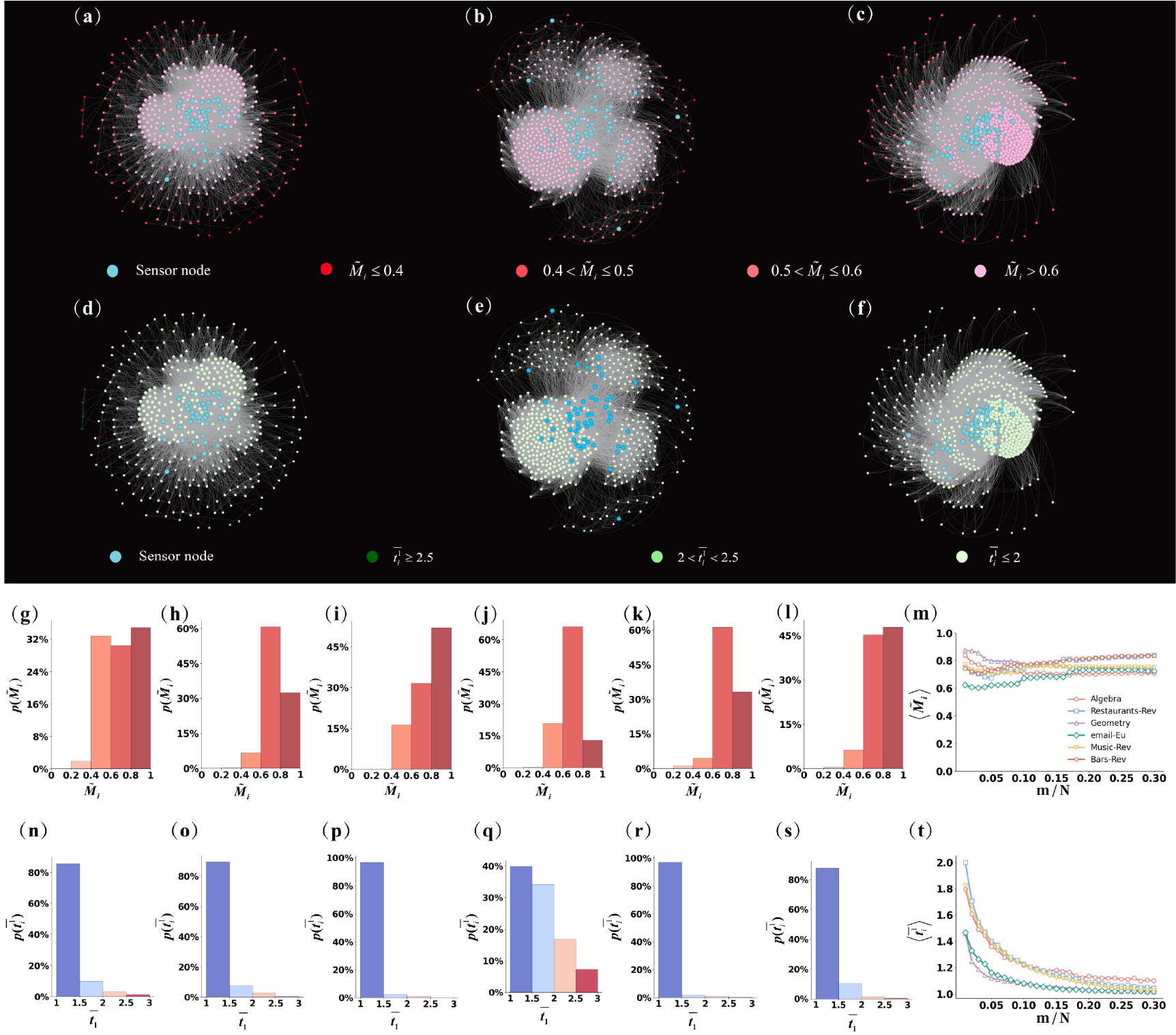}
    \caption{Visualization and quantitative assessment of sensor monitoring coverage across the empirical hypergraphs. (a--c) Node observation intensity ($\tilde{M_i}$) in Algebra, Restaurants-Rev, and Geometry. Sensor nodes are shown in blue, while non-sensor nodes are colored based on their normalized observation intensity: $\tilde{M_i} \leq 0.4$, $0.4 < \tilde{M_i} \leq 0.5$, $0.5 < \tilde{M_i} \leq 0.6$, and $\tilde{M_i} > 0.6$, represented by a gradient from red to pink. Node size corresponds to the $k$-core value, with a larger size indicating a higher $k$-core value. The sensor ratio is set to 10\%. (d--f) Average first-sensor infection time (AFIT, $\bar{t_i^{1}}$) for the same hypergraphs. The AFIT values are visualized using a color gradient from dark to light green, corresponding to $\bar{t_i^{1}} \ge 2.5$, $2 < \bar{t_i^{1}} \le 2.5$, and $\bar{t_i^{1}} \le 2$, respectively. The same sensor set and node sizes as in (a--c) are used, with the sensor ratio fixed at 10\%. (g--l) Distribution of $\tilde{M_i}$ across six empirical hypergraphs under a 10\% sensor ratio. (m) Variation of the average observation intensity $\langle \tilde{M_i}\rangle$ with different sensor ratios in the empirical hypergraphs. (n--s) Distribution of AFIT ($\bar{t_i^{1}}$) across six empirical hypergraphs with a 10\% sensor ratio. (t) Variation of the average AFIT $\langle \bar{t_i^{1}} \rangle$ with sensor ratios in the empirical hypergraphs. All results are obtained using the hypergraph SI spreading model with $\lambda = 0.5$ and averaged over 100 independent Monte Carlo simulations.}
    \label{fig:case1}
\end{figure}

As the deployment of sensors in a hypergraph provides the structural foundation and guidance for the subsequent source detection algorithm, we perform a multidimensional visualization and quantitative evaluation of monitoring coverage across the empirical hypergraphs, whose structural characteristics are summarized in Table~\ref{infor} and described in greater detail in Appendix~\ref{data}. This analysis serves as the motivation for the design of our source detection algorithm presented in the next section, while the effectiveness of our sensor deployment strategy is further validated against state-of-the-art baselines in the experimental section. Our evaluation considers two complementary dimensions:
(1) Node observation intensity, which quantifies the cumulative sensing influence exerted by all deployed sensors and reflects how effectively each node is monitored within the hypergraph; and (2) Average first-sensor infection time, which characterizes the timeliness of detection by measuring how quickly the diffusion process reaches the sensors.

\textbf{Node Observation Intensity.} We define the influence exerted by a sensor $o_j$ on a node $v_i$ as the reciprocal of their shortest-path distance, expressed as $1/d(i,o_j)$. Accordingly, the overall observation intensity of node $v_i$ is computed as the cumulative contribution of all sensors, that is,
\begin{equation}
    M_i = \sum_{o_j \in O} \frac{1}{d(i,o_j)}.
\end{equation}
A larger $M_i$ value indicates that node $v_i$ is more strongly monitored by the deployed sensors. 
To eliminate disparities arising from differences in sensor quantity and network scale among hypergraphs, we normalize $M_i$ as 
\begin{equation}
    \tilde{M_i} = \frac{M_i}{M_{\max}},
\end{equation}
where $M_{\max}$ denotes the maximum observation intensity across all nodes in the hypergraph. 
The spatial distribution of $M_i$ in various empirical hypergraphs is illustrated in Fig.~\ref{fig:case1}(a--c, g--m). 
In Figs.~\ref{fig:case1}(a--c), colors represent four levels of observation intensity in the Algebra, Restaurants-Rev, and Geometry datasets, respectively, where 
$\tilde{M_i} \leq 0.4$, 
$0.4 < \tilde{M_i} \leq 0.5$, 
$0.5 < \tilde{M_i} \leq 0.6$, 
and $\tilde{M_i} > 0.6$ 
are depicted using a gradient from red to pink. 
Blue nodes indicate the sensors (10\% of all nodes), and node size reflects the $k$-core index, with larger nodes corresponding to higher $k$-core values. 
The nodes are spatially arranged according to their $k$-core values, such that highly central nodes are positioned near the core of the hypergraph. 
As shown, the sensors selected by our strategy are typically associated with higher $k$-core values, suggesting that structurally central nodes are more effective for hypergraph monitoring.
Moreover, nodes with higher observation intensity ($\tilde{M_i} > 0.6$) are concentrated in the central region, while those with lower values ($\tilde{M_i} \leq 0.4$) tend to appear at the periphery. 
Fig.~\ref{fig:case1}(g--l) presents the distribution of $\tilde{M_i}$ for six empirical hypergraphs, i.e., Algebra, Restaurants-Rev, Geometry, Email-Eu, Music-Rev, and Bars-Rev, under a 10\% sensor ratio. 
The results show that approximately 70\% of non-sensor nodes achieve $\tilde{M_i} > 0.6$ in each hypergraph, indicating that most nodes are adequately monitored under this configuration. 
Fig.~\ref{fig:case1}(m) further reports the average observation intensity $\langle \tilde{M_i} \rangle$ across different hypergraphs as the sensor ratio varies, revealing that it remains stably around 0.8. 
This demonstrates that the proposed deployment strategy ensures consistently high monitoring coverage across diverse hypergraph structures and sensor ratios.

\textbf{Average First-Sensor Infection Time (AFIT).}
We introduce the average first-sensor infection time to characterize how quickly an infection initiated by a source node $v_i$ propagates to the sensor set in a hypergraph. Specifically, when node $v_i$ is chosen as the infection source, we perform $R = 100$ independent realizations of the SI process with an infection probability $\lambda = 0.5$. In each run, we record $t_i^{1}(r)$, denoting the time step when the \textit{first} sensor node becomes infected. The AFIT of node $v_i$ is then defined as
\begin{equation}
\bar{t_i^{1}} = \frac{1}{R}\sum_{r=1}^{R} t_i^{1}(r),
\end{equation}
where a smaller $\bar{t_i^{1}}$ value implies that node $v_i$ is more likely to trigger early activation of sensors.
Analogous to the normalized observation intensity $\tilde{M_i}$, the distribution of $\bar{t_i^{1}}$ is illustrated in Fig.~\ref{fig:case1}(d–f, n–t). Figures~\ref{fig:case1}(d–f) visualize the AFIT patterns for the Algebra, Restaurants-Rev, and Geometry hypergraphs under a sensor ratio of 10\%. Sensor nodes are highlighted in blue, while other nodes are color-coded according to their AFIT values using a gradient from dark green to light green, corresponding to $\bar{t_i^{1}} \ge 2.5$, $2 < \bar{t_i^{1}} \le 2.5$, and $\bar{t_i^{1}} \le 2$, respectively. Node sizes represent their $k$-core indices, consistent with Fig.~\ref{fig:case1}(a–c).

The results indicate that most nodes display small AFIT values (light green, $\bar{t_i^{1}} \le 2$), whereas only a few peripheral nodes experience delayed infection detection (dark green). The quantitative AFIT distributions for the six empirical hypergraphs with a 10\% sensor ratio are shown in Fig.~\ref{fig:case1}(n–s). With the exception of the Email-Eu hypergraph, over 80\% of nodes exhibit $\bar{t_i^{1}} \le 1.5$, and this fraction nearly reaches 100\% in the Geometry and Music-Rev datasets. The Email-Eu hypergraph, by contrast, demonstrates a noticeably longer average detection time (Fig.~\ref{fig:case1}(q)), although approximately 70\% of its nodes still achieve AFIT values below 2. Overall, these observations confirm that sensors are able to intercept infection signals at an early stage in most scenarios, reflecting the strong early-warning capability of the proposed deployment strategy. Moreover, Fig.~\ref{fig:case1}(t) summarizes the average AFIT across all empirical hypergraphs under varying sensor ratios. As the sensor ratio increases, AFIT consistently decreases and asymptotically approaches 1. Even when the sensor ratio is as low as 0.01, the mean AFIT remains close to 2, underscoring the robustness and efficiency of the sensor placement scheme in achieving rapid detection. 

\subsection{Source Detection Algorithm}
During the spreading process, each sensor node records the time at which it becomes infected and the identity of the node responsible for transmitting the infection to it, referred to as its infector. Leveraging the infection records gathered by the sensors together with the hypergraph structure, we design an algorithm to infer the most likely source of infection. Initially, the set of suspected source nodes, denoted by $\hat{V}$, includes all nodes in the hypergraph, i.e., $\hat{V} = V$. Since this candidate set is typically very large, the source detection module of the proposed SSDH framework operates in two sequential stages. The first stage (candidate screening) performs a coarse filtering to exclude nodes with a low likelihood of being the source, while the second stage (source refinement) further refines the remaining candidates to identify the most probable source node.

\begin{figure}
    \centering
    \includegraphics[width=\linewidth]{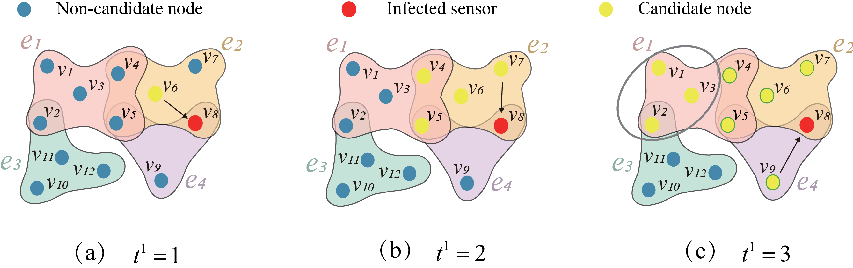}
    \caption{Illustration of the candidate screening process under different earliest sensor infection times ($t^1$). (a) $t^1 = 1$; (b) $t^1 = 2$; and (c) $t^1 = 3$. In each case, node $v_8$ represents the sensor node, while the yellow nodes indicate the corresponding candidate source nodes identified based on the temporal constraint.}
    \label{fig:t_1_case}
\end{figure}

\textbf{Candidate Screening.} This stage efficiently reduces the initial set of potential sources through a hypergraph-based backtracking procedure, complemented by a filtering mechanism to accommodate delayed sensor responses. We first define the set of sensors infected at the earliest infection time $t^1$ as $O^1$, and denote their corresponding infectors as $V_{O^1}$. Since the transmission paths from each infector to its infected sensor are predetermined, they are excluded from further inference. The subsequent analysis therefore focuses on estimating the reachability probability from each candidate source to the nodes in $V_{O^1}$, inferring the infection origin by examining their structural and temporal correlations. According to the value of $t^1$, the screening process is divided into several distinct cases:
\begin{itemize}
    \item When $t^1 = 1$, the sensor node is infected at the first propagation step, implying that its infector corresponds to the initial source of transmission. Hence, the set of suspected source nodes is defined as $\hat{V} = V_{O^1}$. As illustrated in Fig.~\ref{fig:t_1_case}(a), during the first time step $(t = 0 \rightarrow 1)$, node $v_6$ infects sensor node $v_8$, yielding $\hat{V} = \{v_6\}$.
    \item When $t^1 = 2$, according to the hypergraph SI spreading model, the true source node $v_s$ transmits the infection during the first step by randomly selecting a hyperedge. Consequently, the nodes in $V_{O^1}$ must reside in at least one hyperedge that also includes $v_s$. We traverse the hyperedge set $E$ and collect all hyperedges containing $V_{O^1}$ into a candidate set denoted as $E_I$. The corresponding set of suspected source nodes $\hat{V}$ is then obtained by merging all non-sensor nodes from these hyperedges, i.e.,
    \begin{equation}
    \hat{V} = \bigcup_{e \in E_I} e \setminus O.
    \end{equation}
    As illustrated in Fig.~\ref{fig:t_1_case}(b), sensor node $v_8$ is infected by node $v_7$ at the second time step $(t = 1 \rightarrow 2)$. Based on the infection timeline, node $v_7$ must have been infected by the source during the first step. Therefore, within the hyperedge $e_2$ containing $v_7$, all non-sensor nodes except $v_8$ are regarded as potential sources, i.e., $\hat{V} =\{v_4,v_5, v_6, v_7\}$.
    \item When $t^1 > 2$, the infector set $V_{O^1}$ may be separated from the true source node $v_s$ by multiple hops (greater than two), making direct inference infeasible, unlike the previous two cases. To address this, we adopt a backward tracing strategy. Since the infection source must lie within $t^1 - 1$ steps upstream of each infector node $v_j \in V_{O^1}$, a $(t^1 - 1)$-step breadth-first search (BFS) is initiated from every $v_j$, traversing the hypergraph in the reverse direction of the spreading process. All nodes reached during the BFS are collected into the candidate set $V_j$. The intersection of these sets yields the final set of suspected source nodes $\hat{V}$, expressed as:
    \begin{equation}
        \hat{V} = \bigcap_{v_j \in V_{O^1}} V_j \setminus O.
    \end{equation}

For $t^1 > 2$, the candidate set $\hat{V}$ obtained in the previous step may remain large. To further refine $\hat{V}$, we introduce a geometric constraint based on the shortest-path principle. This constraint is motivated by the sensor monitoring characteristics shown in Fig.~\ref{fig:case1}. As observed in Fig.~\ref{fig:case1}(a--c), peripheral nodes in hypergraphs generally exhibit lower connectivity, leading to fewer propagation pathways and longer distances to sensors. Consequently, infections originating from these nodes require more time to reach sensors, as reflected by higher average first-sensor detection times $\bar{t^1}$ (see Fig.~\ref{fig:case1}(d--f), where these nodes typically satisfy $\bar{t^1} > 2$). Therefore, we infer that when $t^1 > 2$, the true source node $v_s$ is likely positioned near the hypergraph periphery, i.e., connected to fewer neighbors and exhibiting high path determinism. Under such conditions, the infection is most likely transmitted through the shortest-path trajectories \cite{brockmann2013hidden,rombach2014core}. To incorporate this structural property and further refine the candidate set $\hat{V}$, we introduce a shortest-path filtering criterion. For each sensor $o_i \in O^1$ infected at time $t^1$ and its corresponding infector $v_i$, a valid candidate source $v_{s'} \in \hat{V}$ must satisfy the following shortest-path condition for at least one such sensor:
\begin{equation}
d(s', o_i) = d(s', i) + 1,
\label{shortest}
\end{equation}
where $d(s', o_i)$ means the distance between $v_{s'}$ and $o_i$, and $d(s', i)$ indicates the distance between $v_{s'}$ and $v_i$.
Nodes that do not meet this criterion are excluded, yielding the final refined candidate set $\hat{V}$. We show an example of $t^1=3$ in Fig.~\ref{fig:t_1_case}(c). Taking the infector node $v_9$ as an example, we perform a backward search of $t^1 - 1 = 2$ steps, resulting in an initial source candidate set $\{v_1, v_2, v_3, v_4, v_5, v_6, v_7\}$. In accordance with Eq.~(\ref{shortest}), the shortest paths from nodes $v_4, v_5, v_6,$ and $v_7$ to the sensor node $v_8$ violate the required condition, since the corresponding paths do not traverse the infector node $v_9$. Hence, these nodes are excluded from the candidate set. Consequently, the refined source candidate set is $\hat{V} = \{v_1, v_2, v_3\}$.
\end{itemize}

\textbf{Source Refinement.} Given the suspected source node set $\hat{V}$, we propose an entropy-based metric to accurately determine the true source node by jointly considering two complementary aspects, i.e., structural proximity and propagation uncertainty. This unified metric captures both the spatial closeness between a candidate source and infected nodes, as well as the rationality of its associated transmission paths. Specifically, for a candidate source node $v_{s'} \in \hat{V}$ and an observation record composed of sensor $o_i$ infected at time $t_i$ and its infector node $v_i$, the metric integrates two components defined as follows:
\begin{itemize}
\item \textbf{Topological distance.} Owing to the designed sensor deployment strategy, the true source node can typically reach the sensors within two propagation steps in most cases (see Fig.~\ref{fig:t_1_case}). Hence, the topological distance between the candidate source $v_{s'}$ and the infected node $v_i \in V_{O^1}$ is expected to be small. We quantify this spatial proximity using the shortest path distance $d(s', i)$ between $v_{s'}$ and $v_i$. A smaller distance indicates stronger topological correlation, implying a higher probability of $v_{s'}$ being the true origin of diffusion.
\item \textbf{Entropy-based transmission.} This component quantifies the uncertainty of the propagation paths from $v_{s'}$ to $v_i$. Lower uncertainty suggests that the propagation from $v_{s'}$ to $v_i$ is more deterministic and thus more plausible.
For a given path $w=(v_0,v_1,\ldots,v_l)$, where $v_0=v_{s'}$ and $v_l=v_i$, the total transmission information cost is computed as the sum of all single-step transition costs:
\begin{equation}
\ell(j,j+1) = \log(1 + k^H_j) - \log(1 + |E_j \cap E_{j+1}|),
\end{equation}
where $k_{j}^H$ denotes the hyperdegree of node $v_j$, and $|E_j \cap E_{j+1}|$ represents the number of shared hyperedges between two consecutive nodes $v_j$ and $v_{j+1}$.
The first term accounts for the uncertainty associated with the size of the propagation potential, while the second term captures the reduction of uncertainty caused by shared hyperedges. The total entropy-based cost along path $w$ is then given by the cumulative sum of $\ell(j,j+1)$. Among all feasible paths from $v_{s'}$ to $v_i$, the minimal cost path defines the path-entropy value:

\begin{equation}
\label{pathscore}
L_{P}(s', i) = \min_{w=(v_0,v_1,\ldots,v_l) \in W(s', i)} \sum_{j=0}^{l-1} \ell(j, j+1),
\end{equation}
where $W(s', i)$ represents the set of paths between node $v_{s'}$ and $v_i$, and a smaller $L_P(s',i)$ indicates lower propagation uncertainty.

\end{itemize}

Combining these two terms, the entropy-based metric is formulated in a weighted manner to yield the final evaluation value, referred to as the source score of node $v_{s'}$, i.e., 
\begin{equation}
    \phi(s') = \sum_{o_i \in O_Y} \frac{1}{t_i} \left[ \beta \cdot d(s', i) + (1-\beta) \cdot L_{P}(s', i) \right],
\end{equation}
where $O_Y$ denotes the set of infected sensors, and $t_i$ and $v_i$ denote the infection time and the corresponding infector node, respectively. The weighting factor $1/t_i$ assigns higher importance to early infection events, which typically exhibit more reliable structural cues. The hyperparameter $\beta \in [0,1]$ balances the contributions of spatial proximity and path uncertainty in the final metric.

\section{Experimental Results}
Our experiments are performed on six synthetic and six empirical hypergraphs, with their key statistical characteristics summarized in Table~\ref{infor}. The synthetic datasets are generated using four representative models, namely the Erd\H{o}s--R\'{e}nyi Hypergraph (ERH) model~\cite{surana2022hypergraph}, the Watts--Strogatz Hypergraph (WSH) model~\cite{watts1998collective}, the Barab\'{a}si--Albert Hypergraph (BAH) model~\cite{feng2024hyper}, and the Hyper-CL (HCL) model~\cite{xie2023efficient}. Detailed generation procedures for these models, as well as descriptions of all empirical datasets, are provided in Appendix~\ref{data}. The empirical hypergraphs encompass a wide spectrum of domains, including social interactions, online review platforms, and communication systems, thereby enabling a comprehensive evaluation of the proposed method under diverse structural configurations and interaction dynamics. To assess source detection performance, we employ two widely used evaluation metrics~\cite{Ali2025acc,zhao2024aed}: accuracy and average error distance (AED). Accuracy quantifies the proportion of correctly identified source nodes, whereas AED represents the mean topological distance between the predicted and actual sources. Hence, higher accuracy and lower AED collectively indicate stronger source localization capability.

To validate the effectiveness of our sensor deployment method, we choose three baselines from the literature, i.e., Greedy Distance Minimization (GDM) \cite{wang2022rapid}, Hierarchical Greedy Coverage (HGC) and Hyperedge Coverage Maximization (HCM). The first two are based on the corresponding simple network of a hypergraph, and the last one is extended from the simple network to a hypergraph by us. Specifically, GDM iteratively selects sensors to minimize the cumulative shortest distance from all nodes to their nearest sensor, HGC prioritizes close-range coverage before extending to greater distances in a layered manner, and HCM selects sensors to maximize coverage of previously uncovered hyperedges. The detailed description of these methods is given in Appendix~\ref{baselinedeploy}.
Subsequently, we benchmark the proposed source detection framework against four representative baselines. The Gradient Maximum Likelihood Algorithm (GMLA) \cite{paluch2018fast} estimates the most probable source via gradient ascent optimization of likelihood functions derived from observed infection times. The Sequential Neighbor Filtering (SNF) method \cite{wang2020locating} infers the source by correlating temporal infection patterns with probabilistic distances computed on a reconstructed cost graph. The Greedy Full-order Neighbor Localization (GFNL) approach \cite{wang2022rapid} integrates multi-faceted indicators, i.e., spatial distance, temporal consistency, and hypergraph-aware reward--penalty mechanisms, to rank source candidates comprehensively. Finally, the Greedy-coverage-based Rapid Source Localization (GRSL) method \cite{wang2023lightweight} achieves computational efficiency through a two-stage procedure combining candidate pre-screening and refined scoring.
Unless specified otherwise, all experiments adopt the default parameter settings: a sensor ratio of 10\%, a final infection ratio $\theta=10\%$, an infection rate $\lambda=0.5$, and a balance coefficient $\beta=0.5$ controlling the trade-off between spatial proximity and path rationality (the sensitive analysis of $\beta=0.5$ is given in Appendix~\ref{parameter}). Further implementation details and evaluation metrics are provided in the Appendix~\ref{baselinedetect}.
\begin{table*}[htp]
    \centering
    \caption{Structural properties of the synthetic and empirical hypergraphs.
Here, $N$ and $M$ denote the numbers of nodes and hyperedges, respectively; $\left\langle k \right\rangle$ represents the average node degree; $\left\langle k^H \right\rangle$ indicates the average hyperdegree; and $\left\langle k^E \right\rangle$ corresponds to the average hyperedge size. The synthetic hypergraphs include ERH-$N$, WSH-$N$, BAH-$N$, and HCL-$\gamma$-$N$, where $N$ denotes the number of nodes and $\gamma$ (in HCL) is the power-law exponent that controls the hyperdegree distribution.}
    \label{infor}
    \begin{tabular}{c@{\hspace{35pt}}c@{\hspace{35pt}}c@{\hspace{35pt}}c@{\hspace{35pt}}c@{\hspace{35pt}}c@{\hspace{35pt}}c}
        \hline
        Hypergraph & $N$ & $M$ & $\left \langle k \right \rangle $ & $\left \langle k^H \right \rangle $ & $\left \langle k^E \right \rangle $  \\ 
        \hline
        ERH-5000 & 5000 & 5000 & 89.18 & 10.00 & 10.00   \\
        WSH-5000 & 5000 & 5000 & 60.61 & 10.00 & 10.00   \\
        BAH-5000 & 5000 & 5000 & 85.97 & 10.00 & 10.00   \\
        HCL-2.0-5000 & 5000 & 5000 & 22.35 & 5.03 & 5.00   \\
        HCL-2.5-5000 & 5000 & 5000 & 25.58 & 5.21 & 5.00   \\
        HCL-3.0-5000 & 5000 & 5000 & 24.23 & 5.10 & 5.00   \\
        Algebra & 423 & 1268 & 78.90 & 19.53 & 6.52  \\
        Restaurants-Rev & 565 & 601 & 79.75 & 8.14 & 7.66   \\
        Geometry & 580 & 1193 & 164.79 & 21.53 & 10.47   \\
        Email-Eu & 998 & 25027 & 58.72 & 85.91 & 3.43   \\
        Music-Rev & 1106 & 694 & 167.88 & 9.49 & 15.13  \\
        Bars-Rev & 1234 & 1194 & 174.3 & 9.62 & 9.94   \\
        \hline
    \end{tabular}
\end{table*}

\textbf{Performance evaluation of the sensor deployment strategy.} For the performance evaluation of the proposed sensor deployment strategy, we adopt two key metrics. The first is the average time to first sensor detection, which measures how quickly the deployed sensors can detect the onset of propagation. A smaller value indicates earlier detection and thus provides more timely information for source localization. As illustrated in Figs.~\ref{fig:case1}(d--f, n--t) and summarized in Table~\ref{Averaget} of the Appendix~\ref{addexp}, our entropy-based deployment strategy achieves significantly shorter detection times compared to baseline methods. Specifically, as shown in Table~\ref{Averaget}, our approach surpasses the second-best algorithm (i.e., HCM) by approximately 1\%--10\% across sensor ratios ranging from 5\% to 15\%. Except for the Email-Eu, when the sensor ratio is 15\%, the average detection time of our method remains below 1.4, demonstrating its superior capability for early information acquisition.
The second evaluation metric involves accuracy and AED under different deployment strategies. As shown in Table~\ref{deploy2}, our method consistently outperforms the second-best algorithm (i.e., HCM) by roughly 1\%--10\% as the sensor ratio varies between 5\% and 15\%, confirming the robustness and effectiveness of our approach. Although the HCM strategy also deploys sensors based on hyperedge structure, it applies uniform selection criteria to all hyperedges and fails to capture structurally critical regions, resulting in an average accuracy gap of about 5\% compared to our method. In contrast, the GDM and HGC algorithms, which are designed for simple networks, are less capable of adapting to higher-order structural dependencies and therefore perform worse. Moreover, our deployment strategy consistently achieves a smaller AED relative to all baselines, further validating its superiority.

\begin{table*}[htp]
    \centering
    \caption{Comparison of accuracy and average error distance (AED) across different sensor deployment strategies for synthetic and empirical hypergraphs. The best results are emphasized in bold to highlight the top-performing method.}
    \label{deploy2}
    
    \resizebox{\textwidth}{!}{%
    \begin{tabular}{l l *{6}{cc}}
        \toprule
         &  & \multicolumn{2}{c}{\textbf{ERH-5000}} & \multicolumn{2}{c}{\textbf{WSH-5000}} & \multicolumn{2}{c}{\textbf{BAH-5000}} & \multicolumn{2}{c}{\textbf{HCL-2.0-5000}} & \multicolumn{2}{c}{\textbf{HCL-2.5-5000}} & \multicolumn{2}{c}{\textbf{HCL-3.0-5000}} \\
        \cmidrule(lr){3-4} \cmidrule(lr){5-6} \cmidrule(lr){7-8} \cmidrule(lr){9-10} \cmidrule(lr){11-12} \cmidrule(lr){13-14}
        \textbf{Methods}&\textbf{Sensor Ratio} & Accuracy & AED & Accuracy & AED & Accuracy & AED & Accuracy & AED & Accuracy & AED & Accuracy & AED \\
        \midrule
        \multirow{3}{*}{GDM} 
        & 5\% &  0.383 & 0.813 & 0.350 & 0.845 & 0.615 & 0.446 & 0.846 & 0.183 & 0.681 & 0.448 & 0.569 & 0.590 \\
        & 10\% &  0.512 & 0.552 & 0.389 & 0.711 & 0.667 & 0.357 & 0.869 & 0.159 & 0.712 & 0.386 & 0.654 & 0.503 \\
        & 15\% & 0.616 & 0.420 & 0.459 & 0.625 & 0.736 & 0.280 & 0.883 & 0.143 & 0.731 & 0.375 & 0.682 & 0.444 \\
        \midrule
        \multirow{3}{*}{HGC} 
        & 5\% & 0.361 & 0.841 & 0.341 & 0.825 & 0.589 & 0.464 & 0.865 & 0.171 & 0.645 & 0.488 & 0.568 & 0.624 \\
        & 10\% & 0.502 & 0.561 & 0.420 & 0.683 & 0.661 & 0.382 & 0.896 & 0.136 & 0.723 & 0.388 & 0.642 & 0.506 \\
        & 15\% & 0.634 & 0.404 & 0.471 & 0.598 & 0.755 & 0.268 & 0.888 & 0.133 & 0.733 & 0.333 & 0.684 & 0.435 \\
        \midrule
        \multirow{3}{*}{HCM} 
        & 5\% & 0.444 & 0.655 & 0.415 & 0.676 & 0.731 & 0.283 & 0.886 & 0.146 & 0.686 & 0.408 & 0.606 & 0.548 \\
        & 10\% & 0.660 & 0.358 & 0.623 & 0.404 & 0.854 & 0.149 & 0.931 & 0.083 & 0.783 & 0.290 & 0.678 & 0.424 \\
        & 15\% & 0.752 & 0.259 & 0.736 & 0.276 & 0.890 & 0.111 & 0.939 & 0.067 & 0.815 & 0.210 & 0.748 & 0.318 \\
        \midrule
        \multirow{3}{*}{Ours} 
         & 5\% & \textbf{0.479}$\uparrow$ & \textbf{0.622}$\downarrow$ & \textbf{0.426}$\uparrow$ & \textbf{0.667}$\downarrow$ & \textbf{0.787}$\uparrow$ & \textbf{0.226}$\downarrow$ & \textbf{0.919}$\uparrow$ & \textbf{0.098}$\downarrow$ & \textbf{0.723}$\uparrow$ & \textbf{0.362}$\downarrow$ & \textbf{0.612}$\uparrow$ & \textbf{0.532}$\downarrow$ \\
         & 10\% & \textbf{0.670}$\uparrow$ & \textbf{0.345}$\downarrow$ & \textbf{0.648}$\uparrow$ & \textbf{0.373}$\downarrow$ & \textbf{0.874}$\uparrow$ & \textbf{0.128}$\downarrow$ & \textbf{0.936}$\uparrow$ & \textbf{0.080}$\downarrow$ & \textbf{0.821}$\uparrow$ & \textbf{0.237}$\downarrow$ & \textbf{0.739}$\uparrow$ & \textbf{0.343}$\downarrow$ \\
         & 15\% & \textbf{0.770}$\uparrow$ & \textbf{0.239}$\downarrow$ & \textbf{0.740}$\uparrow$ & \textbf{0.273}$\downarrow$ & \textbf{0.919}$\uparrow$ & \textbf{0.083}$\downarrow$ & \textbf{0.946}$\uparrow$ & \textbf{0.061}$\downarrow$ & \textbf{0.844}$\uparrow$ & \textbf{0.192}$\downarrow$ & \textbf{0.792}$\uparrow$ & \textbf{0.282}$\downarrow$ \\
         
         \midrule
         &  & \multicolumn{2}{c}{\textbf{Algebra}} & \multicolumn{2}{c}{\textbf{Restaurants-Rev}} & \multicolumn{2}{c}{\textbf{Geometry}} & \multicolumn{2}{c}{\textbf{Email-Eu}} & \multicolumn{2}{c}{\textbf{Music-Rev}} & \multicolumn{2}{c}{\textbf{Bars-Rev}} \\
        \cmidrule(lr){3-4} \cmidrule(lr){5-6} \cmidrule(lr){7-8} \cmidrule(lr){9-10} \cmidrule(lr){11-12} \cmidrule(lr){13-14}
        \textbf{Methods}&\textbf{Sensor Ratio} & Accuracy & AED & Accuracy & AED & Accuracy & AED & Accuracy & AED & Accuracy & AED & Accuracy & AED \\
        
        \midrule
        \multirow{3}{*}{GDM} 
        & 5\% & 0.663 & 0.454 & 0.583 & 0.514 & 0.774 & 0.309 & 0.339 & 1.135 & 0.514 & 0.570 & 0.643 & 0.464 \\
        & 10\% & 0.773 & 0.338 & 0.719 & 0.350 & 0.717 & 0.436 & 0.515 & 0.763 & 0.566 & 0.520 & 0.765 & 0.315 \\
        & 15\% & 0.838 & 0.212 & 0.777 & 0.272 & 0.753 & 0.350 & 0.588 & 0.629 & 0.570 & 0.510 & 0.801 & 0.251 \\
        \midrule
        \multirow{3}{*}{HGC} 
        & 5\% &  0.682 & 0.456 & 0.565 & 0.561 & 0.832 & 0.244 & 0.420 & 0.967 & 0.733 & 0.295 & 0.611 & 0.502 \\
        & 10\% & 0.778 & 0.303 & 0.697 & 0.379 & 0.911 & 0.119 & 0.539 & 0.726 & 0.790 & 0.237 & 0.766 & 0.308 \\
        & 15\% &  0.817 & 0.230& 0.804 & 0.242  & 0.929 & 0.098 & 0.589 & 0.622 & 0.813 & 0.221 & 0.793 & 0.283 \\
        \midrule
        \multirow{3}{*}{HCM} 
        & 5\% & 0.726 & 0.401 & 0.639 & 0.430 & 0.902 & 0.127 & 0.421 & 0.904 & 0.825 & 0.194 & 0.677 & 0.397 \\
        & 10\% & 0.821 & 0.253 & 0.773 & 0.262 & 0.938 & 0.075 & 0.558 & 0.685 & 0.934 & 0.076 & 0.784 & 0.242  \\
        & 15\% & 0.896 & 0.131 & 0.854 & 0.171 & 0.96 & 0.044 & 0.651 & 0.494 & 0.963 & 0.043 & 0.867 & 0.145 \\
        \midrule
        \multirow{3}{*}{Ours} 
         & 5\% & \textbf{0.769}$\uparrow$ & \textbf{0.306}$\downarrow$ & \textbf{0.736}$\uparrow$ & \textbf{0.326}$\downarrow$ & \textbf{0.918}$\uparrow$ & \textbf{0.111}$\downarrow$ & \textbf{0.465}$\uparrow$ & \textbf{0.854}$\downarrow$ & \textbf{0.896}$\uparrow$ & \textbf{0.135}$\downarrow$ & \textbf{0.738}$\uparrow$ & \textbf{0.353}$\downarrow$ \\
         & 10\% & \textbf{0.854}$\uparrow$ & \textbf{0.172}$\downarrow$ & \textbf{0.805}$\uparrow$ & \textbf{0.251}$\downarrow$ & \textbf{0.941}$\uparrow$ & \textbf{0.070}$\downarrow$ & \textbf{0.595}$\uparrow$ & \textbf{0.588}$\downarrow$ & \textbf{0.942}$\uparrow$ & \textbf{0.066}$\downarrow$ & \textbf{0.850}$\uparrow$ & \textbf{0.187}$\downarrow$ \\
         & 15\% & \textbf{0.897}$\uparrow$ & \textbf{0.123}$\downarrow$ & \textbf{0.872}$\uparrow$ & \textbf{0.152}$\downarrow$ & \textbf{0.962}$\uparrow$ & \textbf{0.040}$\downarrow$ & \textbf{0.710}$\uparrow$ & \textbf{0.396}$\downarrow$ & \textbf{0.972}$\uparrow$ & \textbf{0.032}$\downarrow$ & \textbf{0.896}$\uparrow$ & \textbf{0.117}$\downarrow$ \\
        \bottomrule
    \end{tabular}
    } 
\end{table*}

\begin{figure}
    \centering
    \includegraphics[width=\linewidth]{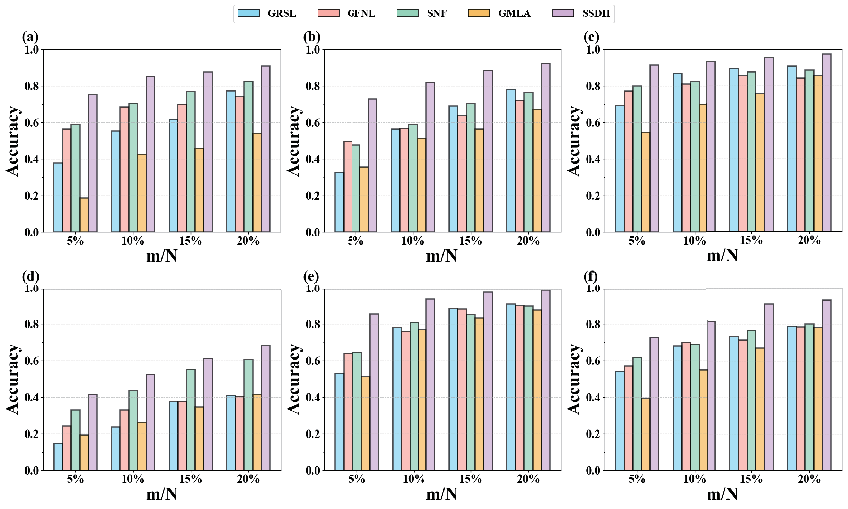}
    \caption{Comparison of source detection performance across different empirical hypergraphs under varying sensor deployment ratios: (a) Algebra; (b) Restaurants-Rev; (c) Geometry; (d) Email-Eu; (e) Music-Rev; (f) Bars-Rev. The performance of the synthetic hypergraphs is shown in Fig.~\ref{fig:sensor1} in the Appendix~\ref{addexp}.}
    \label{fig:sensor2}
\end{figure}

\textbf{Performance evaluation of the source detection method.}  We evaluate the performance of the proposed SSDH algorithm by comparing it with four representative baselines, i.e., GRSL, GFNL, SNF, and GMLA, under default parameter settings across various hypergraphs. Fig.~\ref{fig:sensor2} illustrates the source detection accuracy of all methods under different sensor deployment ratios, while the corresponding results on synthetic hypergraphs are provided in Fig.~\ref{fig:sensor1} of the Appendix~\ref{addexp}. 
The performance evaluated by AED, shown in Figs.~\ref{fig:AEDsensor1} and~\ref{fig:AEDsensor2} of the Appendix~\ref{addexp}, exhibits trends consistent with the accuracy results; therefore, we focus our analysis on accuracy here. As shown in Fig.~\ref{fig:sensor2}, the detection accuracy of all methods improves as the proportion of deployed sensors increases. Notably, SSDH consistently achieves the highest accuracy across all datasets and sensor ratios, maintaining a margin of approximately 5\%--20\% over the best-performing baseline. Even in low-budget scenarios with a 10\% sensor ratio, SSDH outperforms most competing methods operating at 20\%, demonstrating its superior ability to exploit limited observational information efficiently. Among the baseline algorithms, SNF achieves comparatively better results, likely due to its partial adaptation to higher-order interactions, i.e., its correlation-based inference framework can capture certain hyperedge-level propagation dependencies. Nevertheless, SSDH outperforms SNF and all other baselines by effectively fusing a proposed path entropy score with conventional propagation distance, thereby enhancing both the robustness and accuracy of source localization.

\begin{figure}
    \centering
    \includegraphics[width=\linewidth]{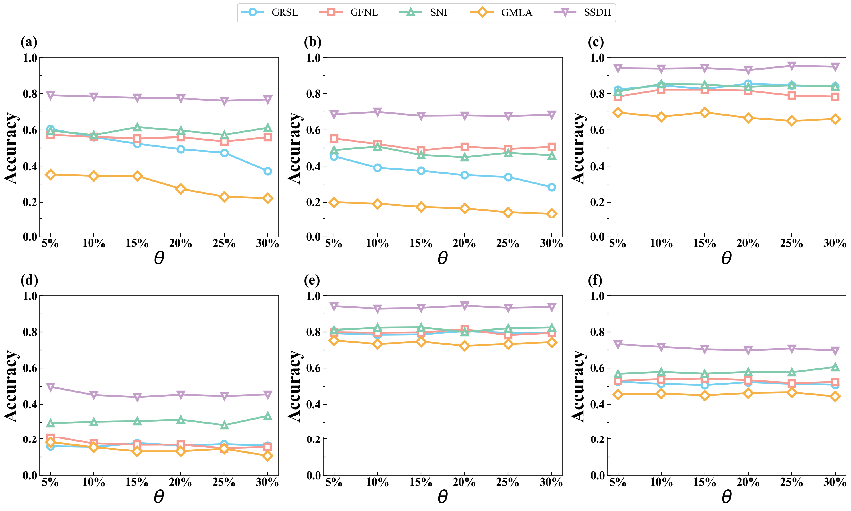}
    \caption{Accuracy of source detection across different methods and hypergraphs under varying final infection ratios $\theta$. Results are shown for six empirical hypergraphs: (a) Algebra, (b) Restaurants-Rev, (c) Geometry, (d) Email-Eu, (e) Music-Rev, and (f) Bars-Rev. The performance of the synthetic hypergraphs is shown in Fig.~\ref{fig:scale1} in the Appendix~\ref{addexp}.}
    \label{fig:scale2}
\end{figure}

\begin{figure}
    \centering
    \includegraphics[width=\linewidth]{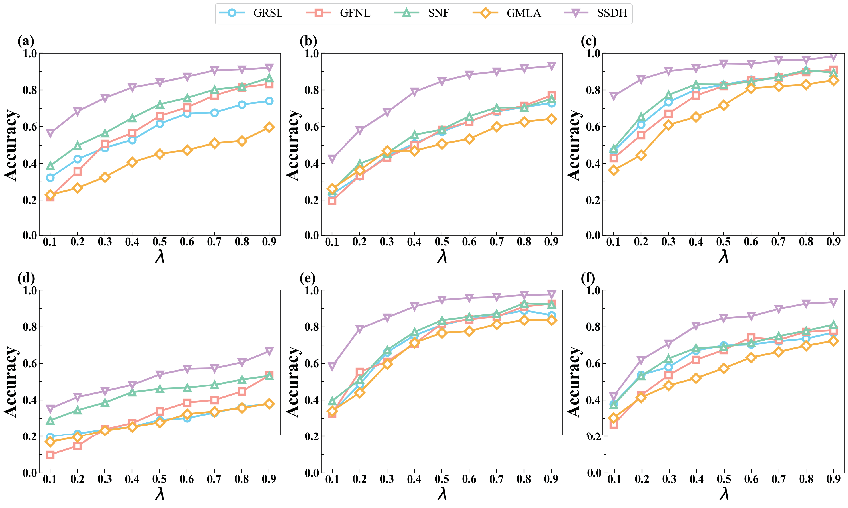}
    \caption{Accuracy of source detection across different methods and empirical hypergraphs under varying propagation probabilities $\lambda$. Results are presented for six empirical hypergraphs: (a) Algebra, (b) Restaurants-Rev, (c) Geometry, (d) Email-Eu, (e) Music-Rev, and (f) Bars-Rev. The corresponding results for synthetic hypergraphs are provided in Fig.~\ref{fig:pro1} of the Appendix~\ref{addexp}.}
    \label{fig:pro2}
\end{figure}

We further conduct a sensitivity analysis on the spreading parameters, namely the final infection ratio $\theta$ and the infection probability $\lambda$, as shown in Figs.~\ref{fig:scale2} and~\ref{fig:pro2} for empirical hypergraphs (and Figs.~\ref{fig:scale1} and~\ref{fig:pro1} for synthetic hypergraphs; Figs.~\ref{fig:AEDscale1}, ~\ref{fig:AEDscale2}, ~\ref{fig:AEDpro1} and ~\ref{fig:AEDpro2} for result of AED in Appendix~\ref{addexp}), to evaluate the robustness of our method. Overall, SSDH demonstrates both stable and superior performance, maintaining an accuracy advantage of approximately 10\%–20\% over the strongest baseline methods under varying $\theta$ values. These findings highlight the robust superiority of SSDH, whereas the baseline algorithms exhibit distinct limitations. Specifically, the SNF algorithm fails to effectively identify source nodes in the early propagation stage due to limited infection observations. In contrast, heuristic methods such as GFNL, GRSL, and GMLA become increasingly vulnerable to noise interference as the spreading process evolves. In comparison, SSDH consistently maintains high detection accuracy across all propagation stages. This robustness primarily stems from its path entropy–based scoring mechanism, which jointly evaluates the likelihood of transmission paths and the reliability of sensor observations. By assigning lower weights to late-infected sensors and filtering redundant or misleading propagation paths, SSDH dynamically adapts to both early and late diffusion scenarios, thereby ensuring reliable, noise-resistant, and resilient source detection performance.

The influence of the spreading probability $\lambda$ on source detection performance is further illustrated in Fig.~\ref{fig:pro2}. The experimental results show that SSDH consistently outperforms all baseline methods across different levels of propagation uncertainty. Remarkably, at higher infection probabilities, SSDH achieves near-perfect accuracy (approaching 1.0) on multiple datasets. While the accuracy of baseline algorithms gradually increases with rising $\lambda$, none are able to surpass SSDH, underscoring its superior robustness and adaptability in diverse diffusion conditions.

\begin{figure}
    \centering
    \includegraphics[width=\linewidth]{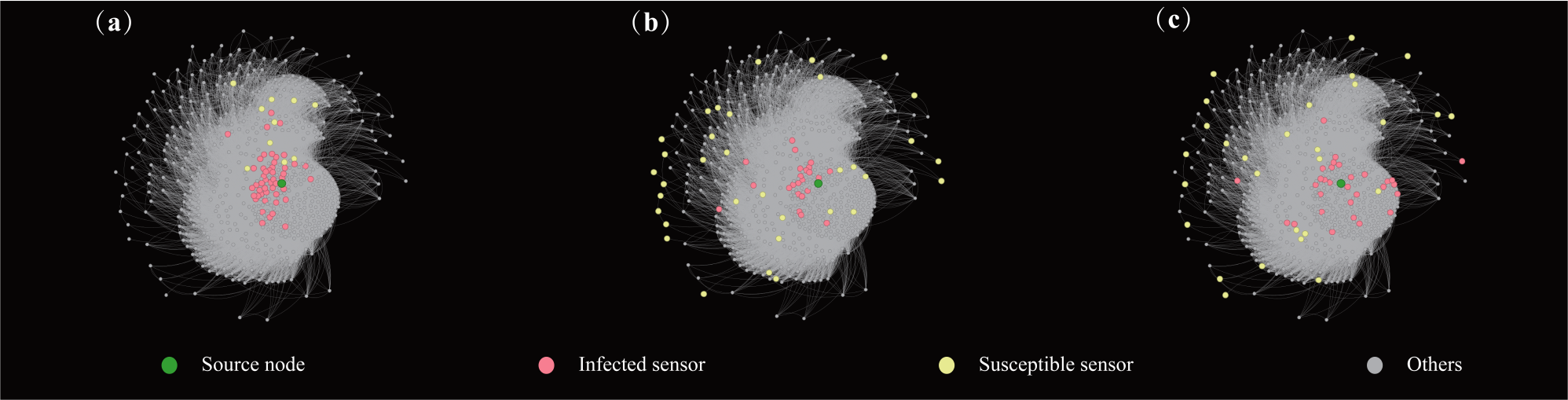}
    \caption{Visualization of source detection results for three algorithms on the Geometry hypergraph. The experiment is conducted with a sensor ratio of 10\%, $\theta = 0.1$, and $\lambda = 0.5$. (a) SSDH; (b) GRSL; (c) SNF.}
    \label{fig:visual2}
\end{figure}

To intuitively illustrate the source detection results, we select the two best-performing baselines, GRSL and SNF, for comparison on the Geometry dataset. In each method, the source node is fixed for visualization, and the spreading parameters are kept identical to ensure a fair comparison. As shown in Fig.~\ref{fig:visual2}, the source node is shown in green, infected sensors in pink, and susceptible sensors in yellow. Compared with the baselines, SSDH yields a larger number of infected sensors that are more spatially concentrated around the true source, thereby facilitating more accurate source identification.
In contrast, the baseline methods produce fewer infected sensors that are more spatially scattered, with some located far from the true source. These distant infections introduce noise and hinder precise localization, further demonstrating the superior reliability of SSDH in capturing the true propagation pattern.

To eliminate the influence of our sensor deployment strategy and evaluate the intrinsic performance of the source detection algorithm, we conduct an additional experiment in which the entropy-based deployment strategy is uniformly applied to all baseline methods. The accuracy of the methods are summarized in Table~\ref{Baselineadd}.
Taking GRSL as an example, Original refers to GRSL using its native sensor deployment method, whereas Optimized denotes the combination of our entropy-based deployment strategy with GRSL’s source detection module. Compared with the original version, GRSL achieves a substantial performance improvement under the optimized configuration, demonstrating the effectiveness of our deployment strategy. This conclusion holds for all optimized baselines.
However, even after optimization, none of the baselines surpass SSDH across different sensor ratios, further confirming that the superiority of SSDH arises not only from the deployment strategy but also from its robust source detection mechanism. Specifically, the sensor deployment component of SSDH ensures the acquisition of high-quality information, while its source detection module maximizes the utilization of that information to achieve the best overall performance.

\begin{table*}[htp]
    \centering
    \caption{Comparison of accuracy for baseline algorithms with and without the entropy-based deployment strategy across different sensor ratios on empirical hypergraphs. The best results are highlighted in bold.}
    \label{Baselineadd}
    
    \resizebox{\textwidth}{!}{%
    \begin{tabular}{l l *{6}{cc}}
        \toprule
         &  & \multicolumn{2}{c}{\textbf{Algebra}} & \multicolumn{2}{c}{\textbf{Restaurants-Rev}} & \multicolumn{2}{c}{\textbf{Geometry}} & \multicolumn{2}{c}{\textbf{Email-Eu}} & \multicolumn{2}{c}{\textbf{Music-Rev}} & \multicolumn{2}{c}{\textbf{Bars-Rev}} \\
        \cmidrule(lr){3-4} \cmidrule(lr){5-6} \cmidrule(lr){7-8} \cmidrule(lr){9-10} \cmidrule(lr){11-12} \cmidrule(lr){13-14}
        \textbf{Methods}&\textbf{Sensors} & Original & Optimized & Original & Optimized & Original & Optimized & Original & Optimized & Original & Optimized & Original & Optimized \\
        \midrule
        \multirow{3}{*}{GRSL} 
        & 5\% &  0.187 & 0.666$\uparrow$ & 0.357 & 0.655$\uparrow$ & 0.546 & 0.869$\uparrow$ & 0.193 & 0.223$\uparrow$ & 0.513 & 0.801$\uparrow$ & 0.393 & 0.694$\uparrow$ \\
        & 10\% &  0.427 & 0.760$\uparrow$ & 0.513 & 0.746$\uparrow$ & 0.700 & 0.878$\uparrow$ & 0.263 & 0.360$\uparrow$ & 0.770 & 0.888$\uparrow$ & 0.550 & 0.793$\uparrow$ \\
        & 15\% & 0.460 & 0.810$\uparrow$ & 0.563 & 0.833$\uparrow$ & 0.756 & 0.898$\uparrow$ & 0.347 & 0.413$\uparrow$ & 0.833 & 0.959$\uparrow$ & 0.673 & 0.861$\uparrow$ \\
        \midrule
        \multirow{3}{*}{GNFL} 
        & 5\% & 0.590 & 0.666$\uparrow$ & 0.477 & 0.655$\uparrow$ & 0.801 & 0.869$\uparrow$ & 0.330 & 0.223$\downarrow$ & 0.646 & 0.801$\uparrow$ & 0.619 & 0.694$\uparrow$ \\
        & 10\% & 0.705 & 0.760$\uparrow$ & 0.590 & 0.746$\uparrow$ & 0.825 & 0.878$\uparrow$ & 0.435 & 0.360$\downarrow$ & 0.811 & 0.897$\uparrow$ & 0.691 & 0.793$\uparrow$ \\
        & 15\% & 0.769 & 0.810$\uparrow$ & 0.704 & 0.833$\uparrow$ & 0.877 & 0.898$\uparrow$ & 0.552 & 0.413$\downarrow$ & 0.852 & 0.959$\uparrow$ & 0.770 & 0.861$\uparrow$ \\
        \midrule
        \multirow{3}{*}{SNF} 
        & 5\% & 0.380 & 0.666$\uparrow$ & 0.328 & 0.656$\uparrow$ & 0.692 & 0.869$\uparrow$ & 0.149 & 0.227$\uparrow$ & 0.531 & 0.801$\uparrow$ & 0.543 & 0.699$\uparrow$ \\
        & 10\% & 0.555 & 0.762$\uparrow$ & 0.563 & 0.748$\uparrow$ & 0.868 & 0.878$\uparrow$ & 0.237 & 0.363$\uparrow$ & 0.781 & 0.897$\uparrow$ & 0.680 & 0.798$\uparrow$ \\
        & 15\% & 0.619 & 0.817$\uparrow$ & 0.690 & 0.839$\uparrow$ & 0.896 & 0.898$\uparrow$ & 0.375 & 0.421$\uparrow$ & 0.884 & 0.967$\uparrow$ & 0.735 & 0.868$\uparrow$ \\
        \midrule
        \multirow{3}{*}{GMLA} 
         & 5\% & 0.563 & 0.673$\uparrow$ & 0.497 & 0.659$\uparrow$ & 0.771 & 0.871$\uparrow$ & 0.243 & 0.233\textbf{$\downarrow$} & 0.639 & 0.805$\uparrow$ & 0.572 & 0.696$\uparrow$ \\
         & 10\% & 0.685 & 0.764$\uparrow$ & 0.567 & 0.753$\uparrow$ & 0.810 & 0.881$\uparrow$ & 0.330 & 0.370$\uparrow$ & 0.761 & 0.900$\uparrow$ & 0.699 & 0.795$\uparrow$ \\
         & 15\% & 0.699 & 0.818$\uparrow$ & 0.638 & 0.836$\uparrow$ & 0.856 & 0.902$\uparrow$ & 0.377 & 0.419$\uparrow$ & 0.882 & 0.959$\uparrow$ & 0.716 & 0.871$\uparrow$ \\
         \midrule
        \multirow{3}{*}{SSDH} 
         & 5\% & \multicolumn{2}{c}{\textbf{0.758}} & \multicolumn{2}{c}{\textbf{0.730}} & \multicolumn{2}{c}{\textbf{0.924}} & \multicolumn{2}{c}{\textbf{0.376}} & \multicolumn{2}{c}{\textbf{0.858}} & \multicolumn{2}{c}{\textbf{0.744}} \\
         & 10\% & \multicolumn{2}{c}{\textbf{0.845}} & \multicolumn{2}{c}{\textbf{0.816}} & \multicolumn{2}{c}{\textbf{0.936}} & \multicolumn{2}{c}{\textbf{0.535}} & \multicolumn{2}{c}{\textbf{0.940}} & \multicolumn{2}{c}{\textbf{0.826}} \\
         & 15\% & \multicolumn{2}{c}{\textbf{0.882}} & \multicolumn{2}{c}{\textbf{0.899}} & \multicolumn{2}{c}{\textbf{0.958}} & \multicolumn{2}{c}{\textbf{0.597}} & \multicolumn{2}{c}{\textbf{0.973}} & \multicolumn{2}{c}{\textbf{0.896}} \\
        \bottomrule
    \end{tabular}
    } 
\end{table*}

\section{Conclusion}
In this work, we systematically investigate the problem of sensor-based source detection on hypergraphs, a task that poses unique challenges due to the combinatorial complexity and higher-order nature of hyperedge interactions. To tackle these challenges, we propose a unified framework termed Sensor-based Source Detection in Hypergraphs (SSDH), which integrates an entropy-based sensor deployment strategy with a hybrid source identification algorithm.

For the sensor deployment phase, we design an entropy-driven strategy grounded in the principle of diminishing marginal returns, ensuring an optimal trade-off between exploration of new structural regions and exploitation of already covered areas. This design enables efficient utilization of limited sensor resources while maintaining high coverage diversity. For the source identification phase, we develop a two-stage hybrid algorithm that first prunes the search space through temporal and geometric constraints, and subsequently refines candidate evaluation using a path entropy–based scoring mechanism that fuses uncertainty quantification with propagation distance. This dual-stage process effectively balances computational efficiency with inference accuracy.

Extensive experiments conducted on both synthetic and empirical hypergraphs demonstrate that SSDH consistently and substantially outperforms all plausible baselines across diverse hypergraph structures and spreading conditions. The results confirm the robustness, adaptability, and interpretability of SSDH, highlighting the crucial role of higher-order interactions in accurately modeling real-world spreading dynamics. The entropy-based deployment ensures the acquisition of high-quality, information-rich observations, while the source detection module maximizes the use of this information through a principled, uncertainty-aware inference process.

Looking ahead, several directions can further enhance this research. Future work could focus on extending SSDH to multi-source and dynamic hypergraph scenarios \cite{Geon2023,Neuh2021}, where interactions evolve over time; integrating lightweight or distributed computation mechanisms to improve scalability for massive real-world networks; and exploring cross-domain applications, such as epidemic tracing, rumor containment, and cyber-attack localization, where the combination of sensor optimization and higher-order reasoning can yield deeper practical insights.

\begin{acknowledgments}
This work was supported by Key Scientific \& Technological Project of XPCC (2025AA018), the National Natural Science Foundation of China (Grant No. 62473123), and the Scientific Research Fund of Zhejiang Provincial Education Department(Y202558115).
\end{acknowledgments}

\section*{DATA AVAILABILITY}
 The data that support the findings of this article are openly available in Refs.~\cite{amburg2020fair,yin2017local,ni2019justifying}.

\appendix
\section{Baseline Algorithms for Sensor Deployment}\label{baselinedeploy}
We compare our entropy-based sensor deployment strategy with three representative baseline methods for sensor placement in hypergraphs. The descriptions of these baseline methods are summarized as follows.

\textbf{Greedy Distance Minimization (GDM)} is a greedy-based sensor deployment approach inspired by the work of Wang et al.~\cite{wang2022rapid}, which aims to iteratively position sensors so as to minimize the total shortest-path distance between all nodes in the hypergraph and their nearest sensors. At each iteration, the algorithm selects the node that yields the greatest reduction in the overall distance sum to the current sensor set, thereby ensuring the most efficient coverage improvement at each step. The shortest distance from a node $v_i$ to the sensor set $O$ is formally defined as:
\begin{equation}
    D(i,O) = \begin{cases}
     \infty, &  O = \emptyset \\
     \min\limits_{o_j\in O} d(i,j), &  O \neq \emptyset
\end{cases}
\end{equation}
If no valid path exists between node $v_i$ and the sensor set $O$, the distance $D(i, O)$ is assigned a value of $2R$, where $R$ denotes the diameter of the corresponding simple network $G$. The overall procedure of the GDM algorithm is summarized as follows:
\begin{enumerate}[label=(\roman*)]
\item Initialize the sensor set $O$ as an empty set.

\item For each candidate node $v_i \in V \setminus O$, compute its distance cost $C(i, O)$, which quantifies the total shortest-distance contribution from all nodes to the updated sensor set $O \cup \{v_i\}$. The cost is defined as:
\begin{equation} C(i,O)=\sum_{v_k \in V\setminus O} \min(d(i, k), D(k, O)), \end{equation}
where $D(k,O)$ denotes the shortest distance from node $v_k$ to the current sensor set. At each iteration, the node with the smallest cost is selected and added to $O$. If multiple nodes achieve the same cost, one of them is chosen at random.

\label{step2}

\item Repeat step~\ref{step2} until the target number of sensors is selected.
\end{enumerate}

\textbf{Hierarchical Greedy Coverage (HGC)}  is a sensor deployment strategy that we have proposed, which considers every neighbour of every order across all nodes. It prioritizes ensuring close-range coverage of hypergraph nodes first, then progressively extends to larger distances. The algorithm employs greedy selection at each coverage level to achieve monitoring of different distances for each node within a limited budget. We define the surveillance range of a node $v_i$ at a specific distance $k$ as:
\begin{equation}
    \Gamma_k(i)=\{v_j\in V\mid d(i,j)=k\}.
\end{equation}
This set contains all nodes that are exactly at a distance $k$ from node $v_i$. The specific steps of the algorithm are as follows:

\begin{enumerate}[label=(\roman*)]
\item Initialize the sensor set $O$ as an empty set and define the set of uncovered nodes as $U = V$.
\item Beginning with a coverage radius $l = 1$, we seek to ensure that each uncovered node has at least one $l$-hop neighbor that is selected as a sensor. To this end, among all non-sensor nodes, we choose the node $v_i$ that maximizes the number of uncovered nodes reachable within distance $l$:
\begin{equation}
    v_{i} = \underset{v_i \in V \setminus O}{\operatorname*{arg\,max}}\, |\Gamma_l(i) \cap U|.
\end{equation}
The chosen node $v_i$ is added to the sensor set $O$, and the uncovered set is updated as $U \leftarrow U \setminus \Gamma_l(i)$.
\label{A2step2}

\item At coverage distance $l$, repeat step~\ref{A2step2} until all nodes are covered, i.e., $U$ becomes empty. If the sensor budget has not yet been fully used, increase the coverage radius to $l \leftarrow l+1$, reset the uncovered set to $U \leftarrow V$, and return to step~\ref{A2step2} to continue the greedy selection.
\label{A2step3}

\item Repeat steps~\ref{A2step2} and~\ref{A2step3} until the required number of sensors has been selected.

\item If no additional nodes can be selected under the above rules, the remaining sensor positions are assigned by choosing nodes uniformly at random.

\end{enumerate}

\textbf{Hyperedge Coverage Maximization (HCM)} is a hypergraph-specific sensor deployment strategy that exploits the fact that contagion in hypergraphs propagates through hyperedges. To enhance monitoring capability, each selected sensor is chosen to cover as many previously uncovered hyperedges as possible. For a node $v_i$, let $E_i$ denote the set of hyperedges incident to $v_i$. The algorithm proceeds as follows:
\begin{enumerate}[label=(\roman*)]
\item Initialize the sensor set $O$ as an empty set, and define the set of uncovered hyperedges as $E' = E$.
\item While the sensor budget remains and there exist hyperedges that are not yet covered ($E' \neq \emptyset$), select the node $v_i \in V \setminus O$ that maximizes the number of uncovered hyperedges it can cover:
\begin{equation}
    v_i = \underset{v_j \in V \setminus O}{\operatorname*{arg\,max}}\, |E_j \cap E'|.
\end{equation}
The selected node is added to the sensor set $O$, and the set of uncovered hyperedges is updated by removing the hyperedges newly covered by $v_i$.
\label{A3step2}

\item Repeat step~\ref{A3step2} until the required number of sensors has been deployed.

\item If no additional nodes satisfy the selection criterion, e.g., all remaining nodes cover only hyperedges already monitored, the remaining sensor slots are filled by selecting nodes uniformly at random.
\end{enumerate}

\section{Baseline Algorithms for source detection}\label{baselinedetect}
In this study, we compare our proposed SSDH algorithm against four baseline methods: Gradient Maximum Likelihood Algorithm (GMLA), Sequential Neighbor Filtering (SNF), Greedy Full-order Neighbor Localizatio (GNFL), and Greedy-coverage-based Rapid Source Localization (GRSL). The details of these baseline algorithms are provided below.

\textbf{Gradient Maximum Likelihood Algorithm (GMLA)}~\cite{paluch2018fast} estimates the infection source in a hypergraph via a likelihood-guided gradient search. Sensor placement follows a random sampling strategy. The algorithm consists of three key stages. First, leveraging the fact that early-infected sensors provide more reliable temporal cues, it selects the $K_0 = 0.5\sqrt{N}$ sensors with the earliest infection timestamps. If any selected sensor is infected at $t = 1$, GMLA immediately identifies its infector as the source. After determining the sensor set, the algorithm computes a likelihood score $\phi(s)$ for each candidate node $v_{s}$. The score is based on the temporal differences between infection times and the corresponding shortest-path distances among selected sensors. Taking the earliest infected sensor $O^1$ as a temporal reference, GMLA constructs two vectors: $\mathbf{d}$ for the observed infection-time differences, and $\boldsymbol{\mu}$ for the expected delays implied by the shortest-path distances from $O^1$ to the other selected sensors. Although infection events occur in discrete steps, the accumulated delay along a path arises from many independent activation attempts. By the Central Limit Theorem, this aggregated delay is approximated by a Gaussian distribution. Under this assumption, $\mathbf{d}$ is modeled as a multivariate normal vector, yielding the likelihood for node $v_{s}$ to be the source:
\begin{equation} \phi({s}) = \frac{\exp\left(-\frac{1}{2}(\mathbf{d}-\boldsymbol{\mu}_{s})^T \boldsymbol{\Lambda}_{s}^{-1}(\mathbf{d}-\boldsymbol{\mu}_{s})\right)}{\sqrt{|(\boldsymbol{\Lambda}_{s})|}}, \end{equation}
where $\boldsymbol{\Lambda}_i$ denotes the covariance matrix. Each entry $(\boldsymbol{\Lambda}_{s})_{jk}$ reflects the number of common edges traversed by the shortest paths from $O^1$ to sensors $o_j$ and $o_k$ in the BFS tree rooted at $v_{s}$. Finally, GMLA adopts a gradient-ascent–like local exploration strategy. Starting from the earliest infected sensor $O^1$, the algorithm evaluates the likelihood scores of all its adjacent nodes and moves to the neighbor with the highest score. This process is repeated in a greedy manner: at each step, the current best candidate expands to its unvisited neighbors, and the candidate set is updated only if a neighbor yields a higher score. The search terminates once no surrounding node offers a better likelihood value, at which point the current node is returned as the estimated source. Finally, GMLA performs a gradient-ascent-style local search. It starts from the earlist infected sensor $O^1$, computes the likelihood scores of its neighbors, selects the neighbor node with the highest score, and iteratively explores the unvisited neighbors of the current best candidate until no neighbor exhibits a higher score.

\textbf{Sequential Neighbor Filtering (SNF)}~\cite{wang2020locating} is a correlation-driven source identification algorithm. It employs a full-order neighbor-coverage strategy for sensor placement, where a greedy procedure iteratively selects nodes that maximize the number of newly covered neighbors across all hop distances. The method is built on the principle that a node’s actual infection time should be positively correlated with its topological distance from the unknown source. To generalize this principle to probabilistic contagion on hypergraphs, we redefine the notion of node-to-node distance through a probabilistic cost metric. Concretely, we construct a cost map in the form of a weighted directed graph, where edge weights are derived from the entropy-based transmission measure introduced in Eq.~(\ref{pathscore}). In this auxiliary graph, the weight of a directed edge from $v_i$ to $v_k$ quantifies the minimal cost required for information to propagate from $v_i$ to $v_k$, computed as follows:
\begin{equation}
L_{P}(i, k) = \min_{w=(v_0,v_1,\ldots,v_l) \in W(i, k)} \sum_{j=0}^{l-1} \ell(j, j + 1),
\end{equation}
where $W(i,k)$ denotes the set of all feasible paths connecting $v_i$ and $v_k$, and $\ell(j,j+1)$ represents the probabilistic transmission cost incurred when moving between consecutive nodes along a path. The distance between two nodes is thus defined as the minimum weighted-path length in this cost map, capturing the least cumulative transmission cost needed for information to propagate from one node to the other. Before constructing the distance and infection-time vectors, SNF performs an early-infection screening step. If any sensor is detected to be infected at $t = 1$, the algorithm immediately assigns the infector of that sensor as the source, bypassing subsequent computations.
For each remaining candidate source node $v_s$, the algorithm constructs a time vector $\boldsymbol{T}_s$ and a distance vector $\boldsymbol{Q}_s$. Specifically, $\boldsymbol{T}_s$ records the infection times of the selected sensors, while $\boldsymbol{Q}_s$ contains the corresponding transmission costs $L_{P}(s,\cdot)$ from the candidate source $v_s$ to each sensor. The sensors are first grouped into layers according to their cost-based distances from $v_s$; within each layer, the earliest-infected sensor is retained to populate the two vectors.
SNF then evaluates each candidate source by computing the Pearson correlation:
\begin{equation}
\phi({s}) = \text{corr}(\boldsymbol{T}_{s}, \boldsymbol{Q}_{s}),
\end{equation}
where $\mathrm{corr}(\cdot,\cdot)$ denotes the Pearson correlation coefficient. The candidate node with the highest score $\phi(s)$ is ultimately identified as the most probable origin of the infection.

\textbf{Greedy Full-order Neighbor Localization (GFNL)}\cite{wang2022rapid} is a comprehensive evaluation framework developed for rapid source identification during the early phase of propagation. It employs the GDM sensor deployment strategy, as detailed in Appendix~A. To extend its source detection component to hypergraphs, we preserve its distance-based evaluation paradigm from simple networks while redesigning its reward–penalty mechanism to better exploit hypergraph structural information. For each candidate source node $v_{s}$, GFNL calculates an aggregated score composed of three multiplicative terms:
\begin{equation}
    \label{baseline3}
    \phi({s}) =  \left( \sum_{o \in O_Y} d'({o,v_s}) \right)^p \cdot \left( \sum_{o \in O_Y} \otimes(t_o, d'({o,v_s})) \right)^q \cdot rp(v_s).
\end{equation}

The first component quantifies the reconstructed effective propagation distance from the candidate source $v_s$ to all infected sensors $o \in O_Y$. Instead of using the direct shortest path to each sensor, the distance is defined as the length of the shortest path from $v_s$ to the infector of the sensor plus one, thereby incorporating directional information and yielding a more accurate measure of spatial proximity. The second component is a time--space proportionality term, where $t_o$ denotes the infection time of sensor $o$. The function $\otimes(t, d') = \max(t/d',\, d'/t)$ penalizes candidates whose propagation time and reconstructed distance are inconsistent. The third component, $rp(v_s)$, is a reward--penalty factor tailored for hypergraphs. For each candidate node $v_s$, the algorithm inspects all hyperedges incident to it: if a hyperedge contains at least one infected sensor, a reward is assigned to $v_s$; if a hyperedge contains sensors but none are infected, a penalty is imposed. The candidate source with the minimal aggregated score is ultimately identified as the most likely origin of the outbreak.


\textbf{Greedy-coverage-based Rapid Source Localization (GRSL)}\cite{wang2023lightweight} builds upon GFNL and is designed as a more lightweight framework for rapid early-stage localization in large-scale hypergraphs. The algorithm accelerates both sensor deployment and source inference. Its deployment scheme adopts an efficient greedy strategy that achieves broad coverage at low computational cost by prioritizing hub and leaf nodes while randomly sampling their full-order neighbors. A key innovation in GRSL lies in the introduction of a pre-screening step for candidate sources. To reduce the initial search space, the algorithm first computes, for each potential source, the cumulative distance to all infected sensors $O_Y$. Only nodes whose distance sum falls below a predefined threshold $l$ are retained in the reliable candidate set $V_l$. The scoring function defined in Eq.~(\ref{baseline3}) is then evaluated solely over these candidates. The three components of this score, i.e., the distance sum, the space--time proportionality term, and the hypergraph-specific reward--penalty factor, are formulated in a manner consistent with GFNL. This two-stage "filter-then-score" design substantially reduces computational overhead while preserving detection accuracy, making GRSL particularly suitable for fast source localization under sparse observations in the early phase of propagation.

\section{Data Description}\label{data}
The experiments in this study are conducted on a collection of twelve hypergraphs, comprising six synthetically generated and six empirical datasets. The synthetic hypergraphs are produced using four representative generative models: the Erdős–Rényi Hypergraph (ERH) \cite{surana2022hypergraph}, the Barabási–Albert Hypergraph (BAH) \cite{feng2024hyper}, the Watts–Strogatz Hypergraph (WSH) \cite{watts1998collective}, and the Hyper-CL (HCL) model \cite{xie2023efficient}.
To ensure transparency and reproducibility, we provide a concise description of the construction procedures for each model in the following section.

\textbf{ERH:} The ERH model generates a $k^E$-uniform hypergraph by sampling hyperedges uniformly at random. For a given node set of size $N$ and a target number of hyperedges $M$, the procedure iteratively draws a subset of $k^E$ distinct nodes to form a candidate hyperedge. If this subset has not appeared previously, it is added to the hyperedge set; otherwise, a new sample is drawn. The sampling continues until exactly $M$ unique hyperedges are obtained. This construction yields a random $k^E$-uniform hypergraph without structural bias.

\textbf{WSH:} The WSH model constructs a $k^E$-uniform hypergraph that exhibits small-world characteristics. The process starts by placing $N$ nodes on a ring and forming a regular structure in which each node participates in a hyperedge with its $k^E - 1$ nearest neighbors, yielding a deterministic $k^E$-ring hypergraph. A rewiring procedure is then applied: for each hyperedge $e$, a candidate hyperedge $e'$ is created by sampling $k^E$ nodes uniformly at random. If $e'$ does not already exist in the hypergraph, the original hyperedge $e$ is replaced by $e'$ with probability $p$. This process is repeated for all hyperedges.

\textbf{SFH:} The SFH model produces $k^E$-uniform hypergraphs in which node degrees follow a power-law distribution, $p(k)\sim k^{-\gamma}$, with $\gamma$ controlling the heaviness of the tail. Given the desired numbers of nodes $N$ and hyperedges $M$, a target degree sequence is first sampled from the prescribed distribution. Each node $v_i$ is then assigned a selection probability $p_i$ proportional to its target degree. Hyperedges are generated sequentially: starting from an empty set $e$, nodes are sampled according to ${p_i}$ and added to $e$ until $|e| = k^E$. If the resulting hyperedge has not appeared previously, it is included in the hypergraph; otherwise, a new attempt is made. Repeating this procedure until $M$ distinct hyperedges are constructed yields a scale-free, $k^E$-uniform hypergraph.

\textbf{HCL:} The HCL model generates hypergraphs with heterogeneous hyperedge sizes. Given the number of nodes $N$ and hyperedges $M$, a node hyperdegree sequence $\{k^H_1,\dots,k^H_N\}$ is first sampled from a power-law distribution, $p(k^H)\sim (k^H)^{-\gamma}$, where $\gamma$ controls the degree heterogeneity. The hyperedge size sequence $\{k^E_1,\dots,k^E_M\}$ is independently drawn from a uniform distribution with an upper bound of 10, allowing non-uniform hyperedge cardinalities. For each hyperedge $e_j$ of target size $k^E_j$, nodes are chosen sequentially with probability $ k^H_i / \sum_{t=1}^N k^H_t$ and added to $e_j$ until its size constraint is met. This procedure is repeated for all $M$ hyperedges.

The six real-world hypergraphs used in this study are described as follows:
\begin{itemize}
    \item \textbf{Algebra} \cite{amburg2020fair}: This dataset is derived from user activity on the MathOverflow platform, focusing specifically on posts tagged with algebra. Each node corresponds to a distinct user, while each hyperedge aggregates all users who participated in the same discussion thread, i.e., whether by posing a question, providing an answer, or contributing a comment.
    \item \textbf{Restaurants-Rev} \cite{amburg2020fair}: This dataset is constructed from Yelp reviews of restaurants located in Madison, Wisconsin. Nodes correspond to individual users, while each hyperedge groups together all users who reviewed establishments belonging to the same restaurant sub-category within a defined time window.
    \item \textbf{Geometry} \cite{amburg2020fair}: Similar in construction to the Algebra dataset, this hypergraph captures user activity on MathOverflow, but restricted to posts tagged with geometry. Each node represents a distinct user, and each hyperedge aggregates all users who contributed to the same discussion thread, by asking, answering, or commenting, thereby encoding higher-order collaborative interactions within the geometry-focused community.
    \item \textbf{Email-Eu} \cite{yin2017local}: This dataset represents internal email communication within a European research institution. Each node corresponds to an individual email address, and each hyperedge encompasses the sender together with all recipients of a single email, thereby capturing the higher-order structure of group communication events.
    \item \textbf{Music-Rev} \cite{ni2019justifying}: This dataset captures the interaction patterns of Amazon reviewers within specific music genres. Nodes correspond to individual reviewers, while each hyperedge groups together all users who reviewed items in a particular blues sub-category during the same month. 
    \item \textbf{Bars-Rev} \cite{amburg2020fair}: It is derived from Yelp reviews of bars located in Las Vegas, Nevada. Nodes represent individual users, and each hyperedge aggregates all users who reviewed establishments within the same bar sub-category over a shared time window.
\end{itemize}

\section{Experimental Analysis}\label{addexp}
To assess the effectiveness of our sensor deployment strategy, we evaluate the average time at which the first sensor becomes infected across the hypergraph datasets. The results, summarized in Table~\ref{Averaget}, demonstrate that our method consistently triggers sensor activation at earlier stages of the spreading process. This advantage suggests that the selected sensors are well positioned to intercept propagation paths promptly and that the deployment strategy successfully identifies structurally influential regions within the hypergraph. Although HCM yields competitive results, its performance remains slightly inferior to ours. This difference stems from a key design distinction: our method explicitly leverages the benefit of placing additional sensors near previously deployed ones to strengthen monitoring in structurally critical areas, whereas HCM prioritizes hyperedge coverage without explicitly accounting for such local reinforcement. As a result, HCM may overlook regions where redundancy enhances early detection. By comparison, the GDM and HGC baselines, both designed on the corresponding simple network rather than the hypergraph itself, struggle to capture the richer higher-order interaction patterns inherent in hypergraphs. Their inability to exploit these multiway relationships leads to noticeably weaker early-detection performance relative to both HCM and our proposed strategy.

\begin{table*}[htp]
\centering
\caption{Average first-sensor infection time achieved by different sensor deployment strategies.}
\label{Averaget}
\resizebox{\textwidth}{!}{%
\begin{tabular}{l l c c c c c c c c c c c c}
\toprule
\multicolumn{2}{c}{} & 
\multicolumn{6}{c}{\textbf{Synthetic Hypergraphs}} & 
\multicolumn{6}{c}{\textbf{Empirical Hypergraphs}} \\
\cmidrule(lr){3-8} \cmidrule(lr){9-14}
 & & 
\textbf{ERH-5000} & \textbf{WSH-5000} & \textbf{BAH-5000} & \textbf{HCL-2.0} & \textbf{HCL-2.5} & \textbf{HCL-3.0} & 
\textbf{Algebra} & \textbf{Restaurants} & \textbf{Geometry} & \textbf{Email-Eu} & \textbf{Music-Rev} & \textbf{Bars-Rev} \\
\midrule

\multirow{3}{*}{GDM} 
 & 5\% & 1.9250 & 1.9980 & 1.5920 & 1.2990 & 1.6270 & 1.7660 & 1.5600 & 1.5960 & 1.2810 & 2.3700 & 1.3680 & 1.4850 \\
 & 10\% & 1.6890 & 1.8470 & 1.4390 & 1.2700 & 1.4900 & 1.6080 & 1.3980 & 1.4610 & 1.2330 & 2.0650 & 1.2540 & 1.3590 \\
 & 15\% & 1.4900 & 1.7480 & 1.3650 & 1.2480 & 1.4480 & 1.5870 & 1.3060 & 1.3110 & 1.1730 & 1.9680 & 1.1230 & 1.2870 \\
\midrule

\multirow{3}{*}{HGC} 
 & 5\% & 1.9090 & 1.9730 & 1.5700 & 1.2970 & 1.6030 & 1.7820 & 1.5040 & 1.6380 & 1.2450 & 2.2140 & 1.3170 & 1.5330 \\
 & 10\% & 1.6820 & 1.7690 & 1.4790 & 1.2210 & 1.5170 & 1.5880 & 1.4200 & 1.4440 & 1.1380 & 2.0520 & 1.2570 & 1.3740 \\
 & 15\% & 1.5000 & 1.7080 & 1.3590 & 1.2430 & 1.4290 & 1.5090 & 1.3250 & 1.3490 & 1.1040 & 1.9800 & 1.2380 & 1.3020 \\
\midrule

\multirow{3}{*}{HCM} 
 & 5\% & 1.7580 & 1.8340 & 1.3460 & 1.2600 & 1.5560 & 1.6810 & 1.5000 & 1.4980 & 1.1650 & 2.1990 & 1.2250 & 1.5010 \\
 & 10\% & 1.4790 & 1.4810 & 1.2150 & 1.1710 & 1.3410 & 1.5030 & 1.3220 & 1.3670 & 1.0960 & 1.9820 & 1.0880 & 1.2830 \\
 & 15\% & 1.3370 & 1.3440 & 1.1190 & 1.1670 & 1.3470 & 1.4360 & 1.2730 & 1.2290 & 1.0700 & 1.8820 & 1.0530 & 1.1890 \\
\midrule

\multirow{3}{*}{Ours} 
 & 5\% & \textbf{1.7430} & \textbf{1.8010} & \textbf{1.3050} & \textbf{1.2510} & \textbf{1.5120} & \textbf{1.6800} & \textbf{1.4530} & \textbf{1.4420} & \textbf{1.1640} & \textbf{2.1830} & \textbf{1.2040} & \textbf{1.4020} \\
 & 10\% & \textbf{1.4760} & \textbf{1.4730} & \textbf{1.1930} & \textbf{1.1620} & \textbf{1.3240} & \textbf{1.4850} & \textbf{1.3160} & \textbf{1.2770} & \textbf{1.0810} & \textbf{1.9650} & \textbf{1.0620} & \textbf{1.2260} \\
 & 15\% & \textbf{1.3270} & \textbf{1.3370} & \textbf{1.1020} & \textbf{1.1650} & \textbf{1.2400} & \textbf{1.3770} & \textbf{1.2900} & \textbf{1.2020} & \textbf{1.0650} & \textbf{1.8450} & \textbf{1.0500} & \textbf{1.1660} \\
\bottomrule
\end{tabular}%
}
\end{table*}

\begin{figure}
    \centering
    \includegraphics[width=0.9\linewidth]{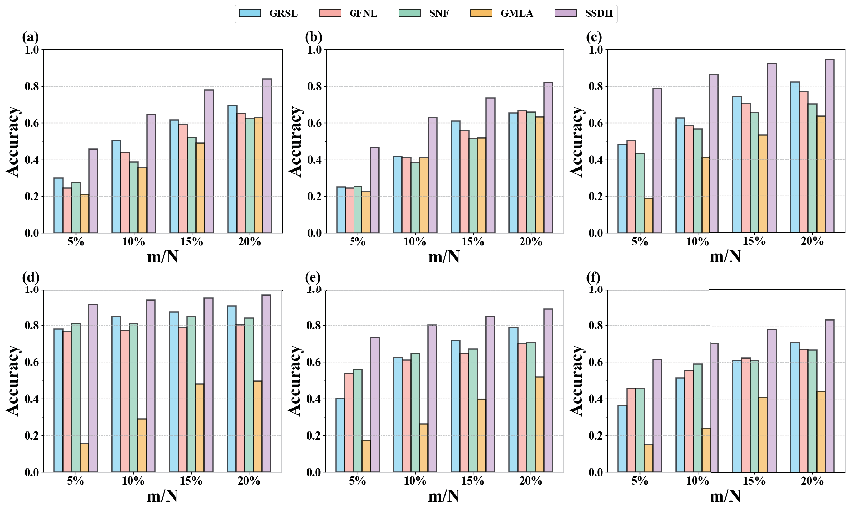}
    \caption{Comparison of source detection performance under varying sensor deployment ratios ($m/N$) across different synthetic hypergraphs: (a) ERH-5000; (b) WSH-5000; (c) BAH-5000; (d) HCL-2.0-5000; (e) HCL-2.5-5000; (f) HCL-3.0-5000.}
    \label{fig:sensor1}
\end{figure}

\begin{figure}
    \centering
    \includegraphics[width=0.9\linewidth]{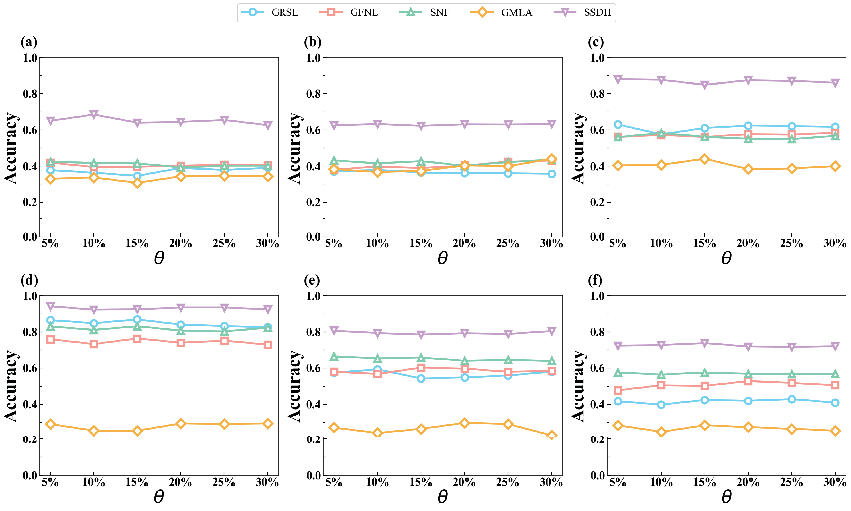}
    \caption{Accuracy of source detection across different methods and hypergraphs under varying final infection ratios $\theta$. Results are shown for six synthetic hypergraphs: (a)ERH-5000 hypergraph; (b)WSH-5000; (c)BAH-5000; (d)HCL-2.0-5000; (e)HCL-2.5-5000; (f)HCL-3.0-5000.}
    \label{fig:scale1}
\end{figure}

\begin{figure}
    \centering
    \includegraphics[width=0.9\linewidth]{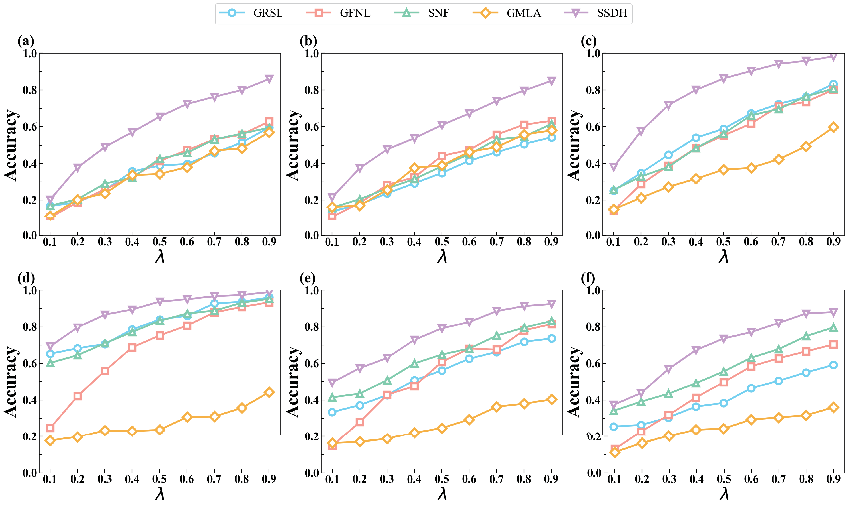}
    \caption{Accuracy of source detection compared across the different methods and synthetic hypergraphs in different propagation probability $\lambda$. (a)ERH-5000 hypergraph; (b)WSH-5000 hypergraph; (c)BAH-5000 hypergraph; (d)HCL-2.0-5000 hypergraph; (e)HCL-2.5-5000 hypergraph; (f)HCL-3.0-5000 hypergraph.}
    \label{fig:pro1}
\end{figure}

\begin{figure}
    \centering
    \includegraphics[width=0.8\linewidth]{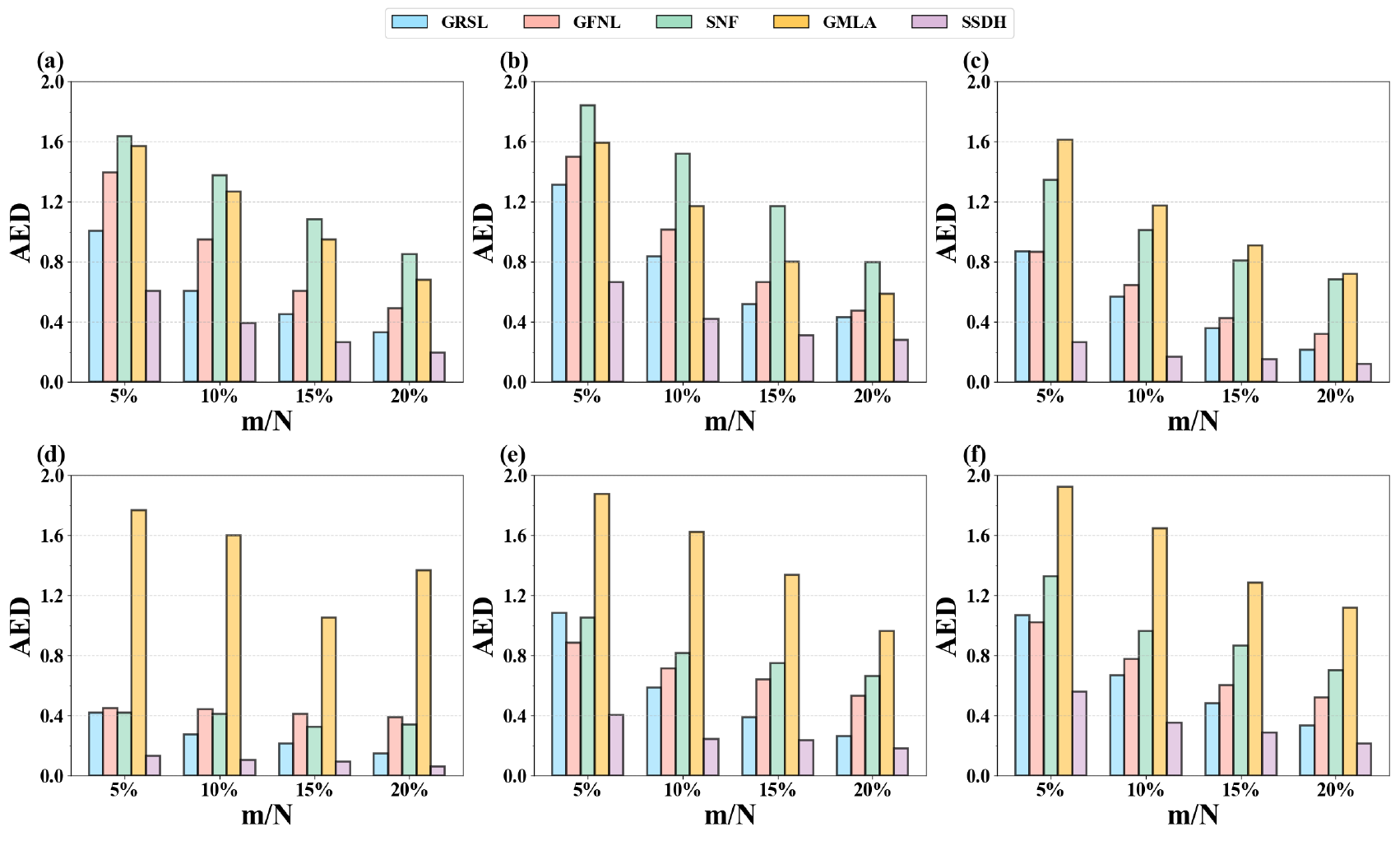}
    \caption{Comparison of average error distance across different synthetic hypergraphs under varying sensor deployment ratios: (a)ERH-5000; (b)WSH-5000; (c)BAH-5000; (d)HCL-2.0-5000; (e)HCL-2.5-5000; (f)HCL-3.0-5000 hypergraph.}
    \label{fig:AEDsensor1}
\end{figure}

\begin{figure}
    \centering
    \includegraphics[width=0.9\linewidth]{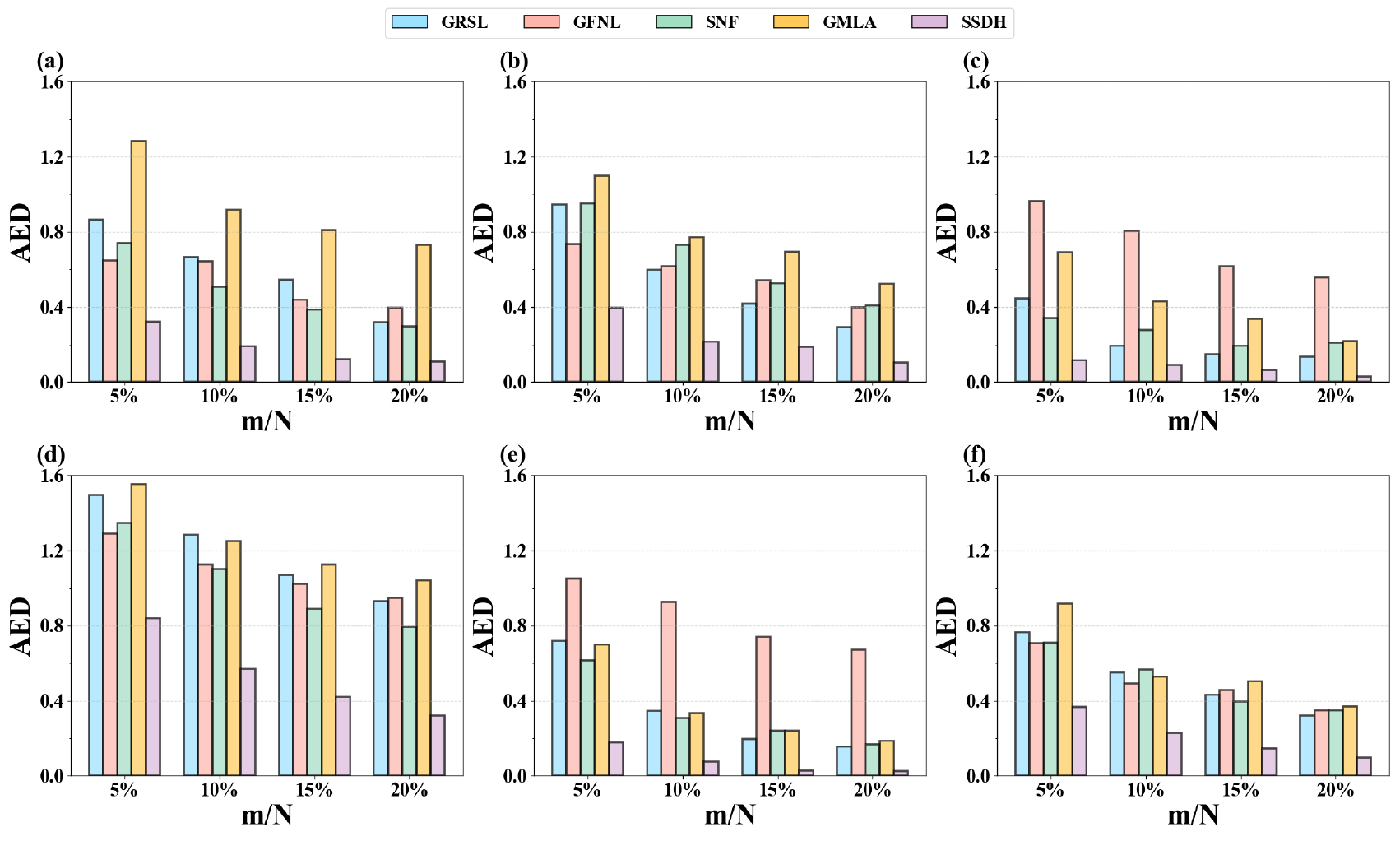}
    \caption{Comparison of average error distance across different empirical hypergraphs under varying sensor deployment ratios: (a) Algebra; (b) Restaurants-Rev; (c) Geometry; (d) Email-Eu; (e) Music-Rev; (f) Bars-Rev.}
    \label{fig:AEDsensor2}
\end{figure}

\begin{figure}
    \centering
    \includegraphics[width=0.9\linewidth]{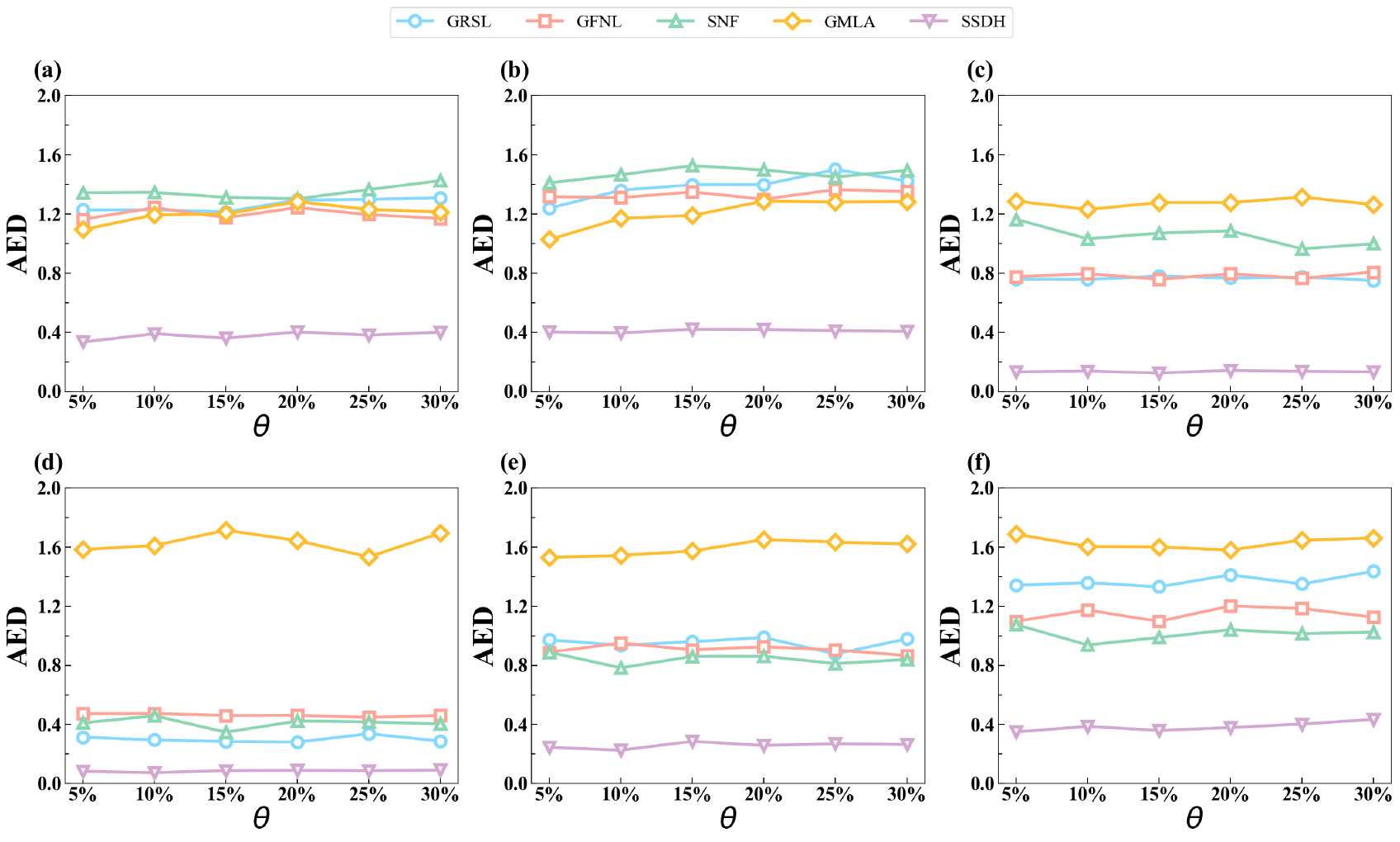}
    \caption{Average error distance of source detection across different methods and hypergraphs under varying final infection ratios $\theta$. Results are shown for six synthetic hypergraphs: (a)ERH-5000 hypergraph; (b)WSH-5000; (c)BAH-5000; (d)HCL-2.0-5000; (e)HCL-2.5-5000; (f)HCL-3.0-5000.}
    \label{fig:AEDscale1}
\end{figure}

\begin{figure}
    \centering
    \includegraphics[width=0.9\linewidth]{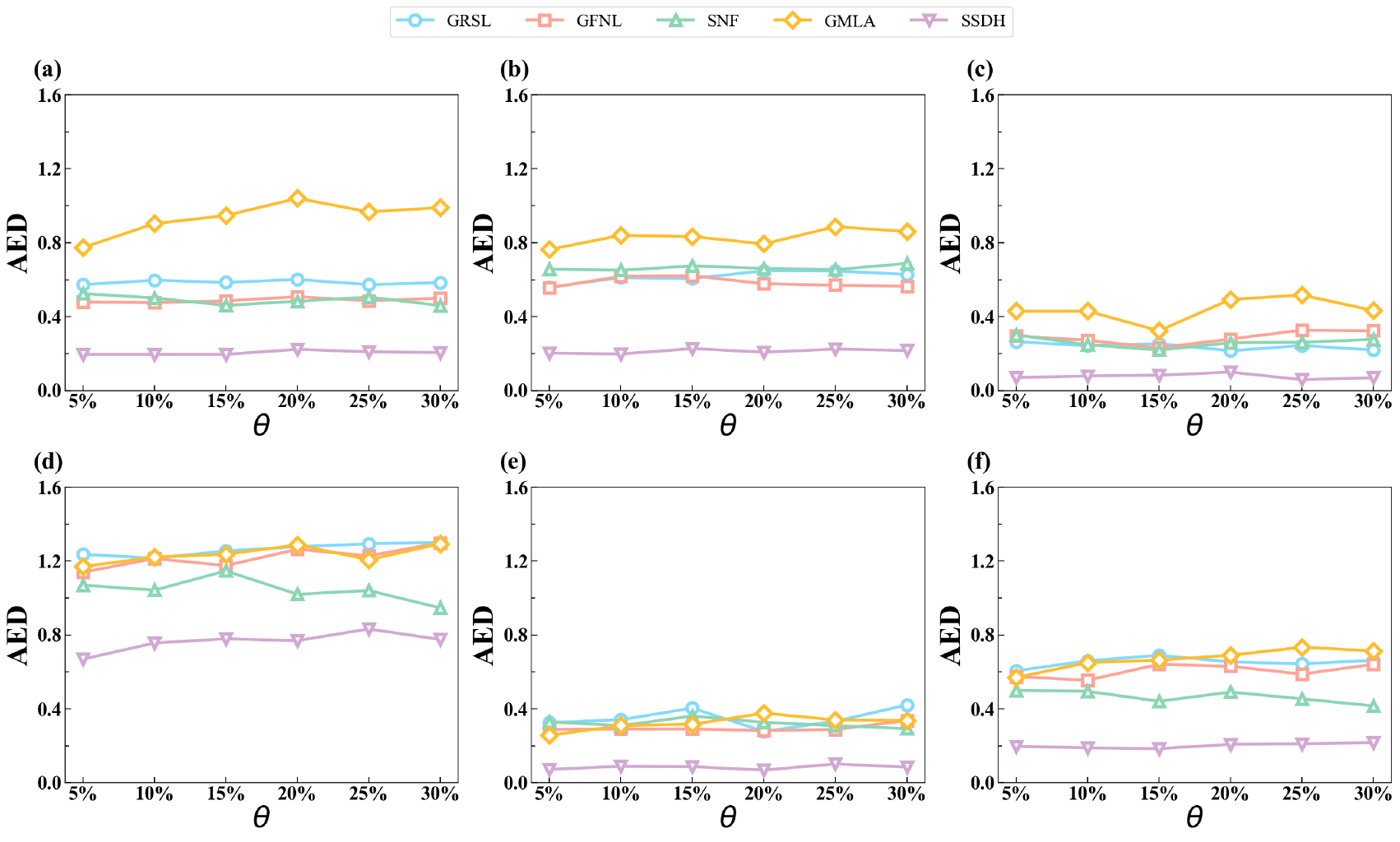}
    \caption{Average error distance of source detection across different methods and hypergraphs under varying final infection ratios $\theta$. Results are shown for six empirical hypergraphs: (a) Algebra, (b) Restaurants-Rev, (c) Geometry, (d) Email-Eu, (e) Music-Rev, and (f) Bars-Rev.}
    \label{fig:AEDscale2}
\end{figure}

\begin{figure}
    \centering
    \includegraphics[width=0.9\linewidth]{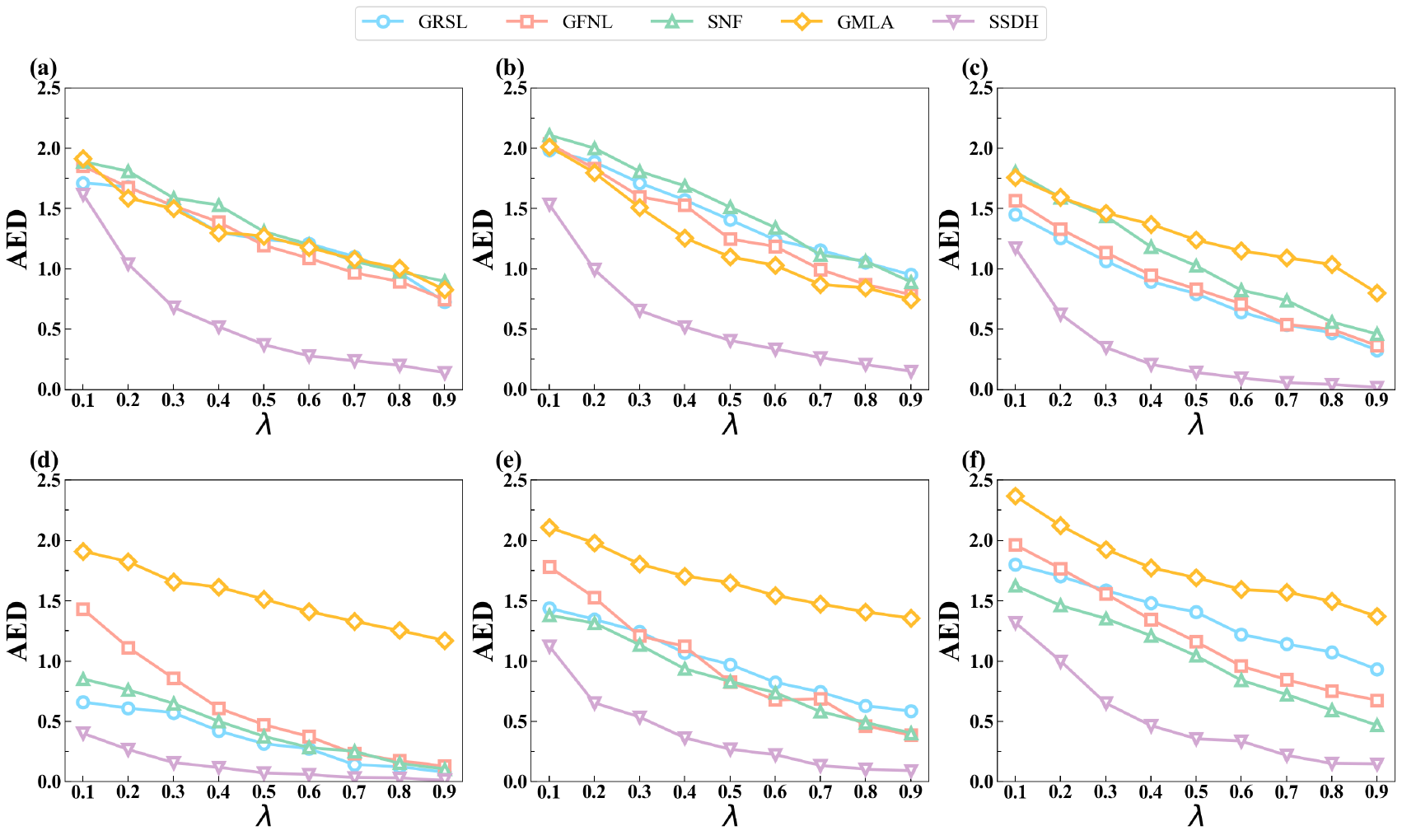}
    \caption{Average error distance of source detection compared across the different methods and synthetic hypergraphs in different propagation probability $\lambda$. Results are shown for six synthetic hypergraphs: (a)ERH-5000 hypergraph; (b)WSH-5000 hypergraph; (c)BAH-5000 hypergraph; (d)HCL-2.0-5000 hypergraph; (e)HCL-2.5-5000 hypergraph; (f)HCL-3.0-5000 hypergraph.}
    \label{fig:AEDpro1}
\end{figure}

\begin{figure}
    \centering
    \includegraphics[width=0.9\linewidth]{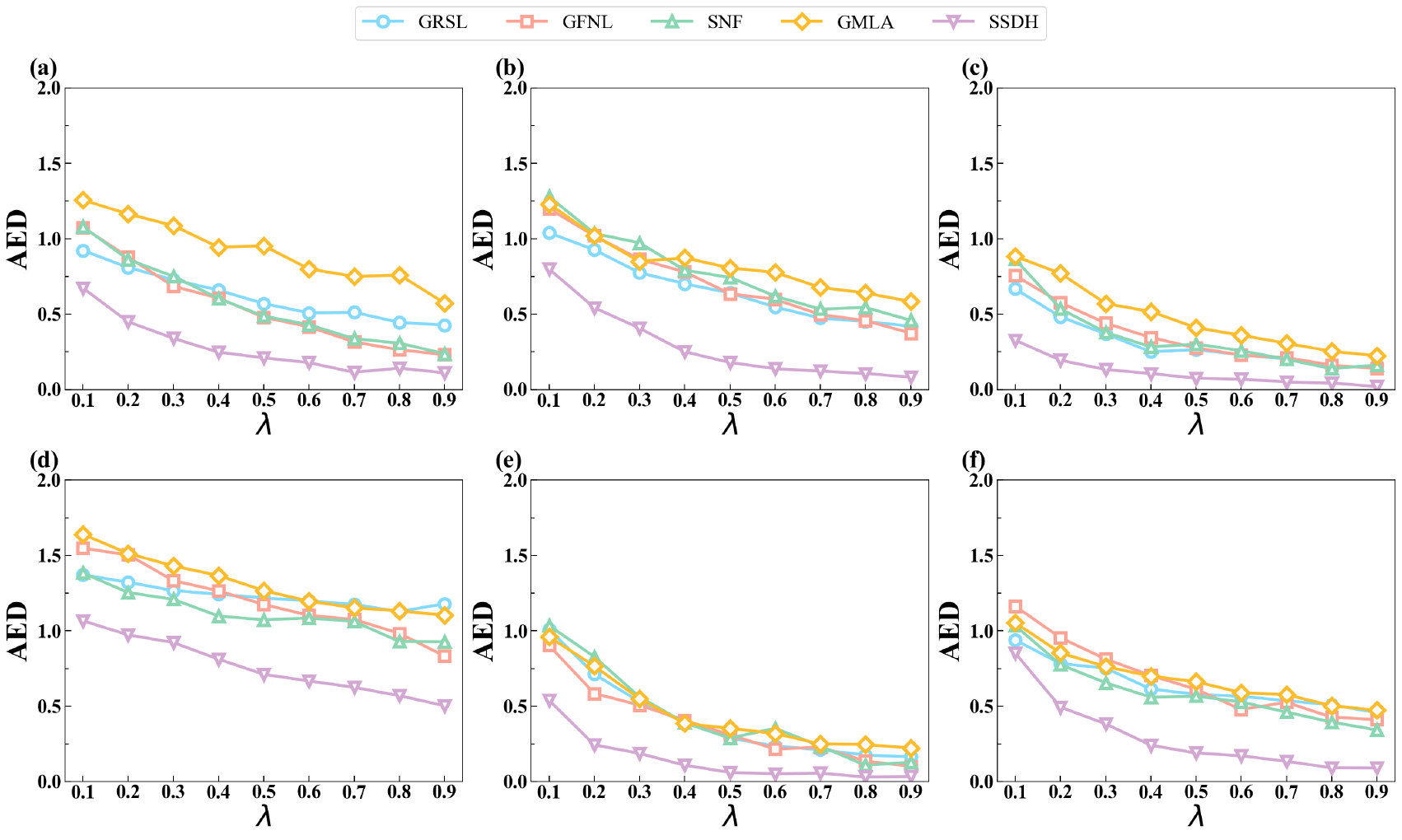}
    \caption{Average error distance of source detection across different methods and empirical hypergraphs under varying propagation probabilities $\lambda$. Results are presented for six empirical hypergraphs: (a) Algebra, (b) Restaurants-Rev, (c) Geometry, (d) Email-Eu, (e) Music-Rev, and (f) Bars-Rev.}
    \label{fig:AEDpro2}
\end{figure}

Our proposed sensor deployment strategy for hypergraphs exhibits consistently strong performance across a wide range of datasets, enabling the early capture of propagation signals and providing critical input for the subsequent source detection task. In addition, we report the source detection accuracy on synthetic hypergraphs as a function of the sensor ratio, propagation scale, and propagation probability, as illustrated in Figs.~\ref{fig:sensor1}, \ref{fig:scale1}, and \ref{fig:pro1}, respectively. These results mirror the trends observed for empirical hypergraphs in the main text (Figs.~\ref{fig:sensor2}, \ref{fig:scale2}, \ref{fig:pro2}), reinforcing that our method consistently surpasses all baseline algorithms under diverse conditions. Because GMLA relies on random sensor placement, it performs on par with the baseline methods only in WSH hypergraphs, where nodal heterogeneity is minimal. In all other cases, its performance is markedly inferior. By comparison, SNF benefits from a more sophisticated distance metric that captures higher-order structural information, yielding improved accuracy across most hypergraphs. For hypergraphs generated by the HCL model, we observe a gradual reduction in accuracy as the power-law exponent $\gamma$ increases. Nevertheless, even under this challenging setting, our SSDH algorithm maintains a clear and stable performance advantage over all competitors. In addition to the accuracy results presented in the main text, we further examine the effects of sensor ratio, propagation scale, and propagation probability using the average error distance (AED). The corresponding comparisons are shown in Figs.~\ref{fig:AEDsensor1}, \ref{fig:AEDsensor2}, \ref{fig:AEDscale1}, \ref{fig:AEDscale2}, \ref{fig:AEDpro1}, and \ref{fig:AEDpro2}. Across all settings, our SSDH algorithm achieves the lowest AED, further confirming its superior source detection capability under diverse propagation and sensing conditions.

\section{Parameter Analysis}\label{parameter}
\begin{figure}
    \centering
    \includegraphics[width=0.9\linewidth]{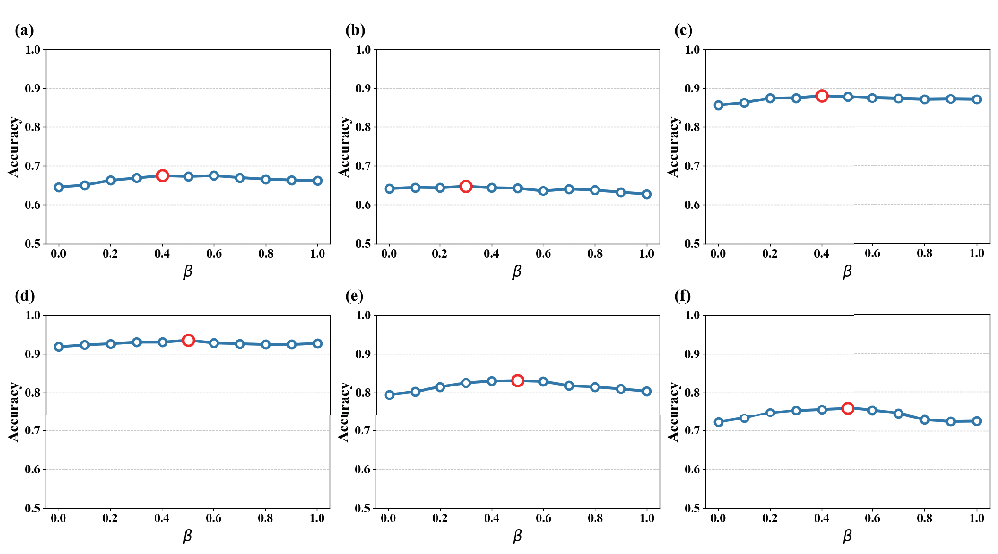}
    \caption{Accuracy of source detection under different values of the parameter $\beta$ and hypergraphs. The red dot indicates the maximum accuracy value within each network. Results are shown for six synthetic hypergraphs: (a)ERH-5000 hypergraph; (b)WSH-5000; (c)BAH-5000; (d)HCL-2.0-5000; (e)HCL-2.5-5000; (f)HCL-3.0-5000.}
    \label{fig:parameter1}
\end{figure}

\begin{figure}
    \centering
    \includegraphics[width=0.9\linewidth]{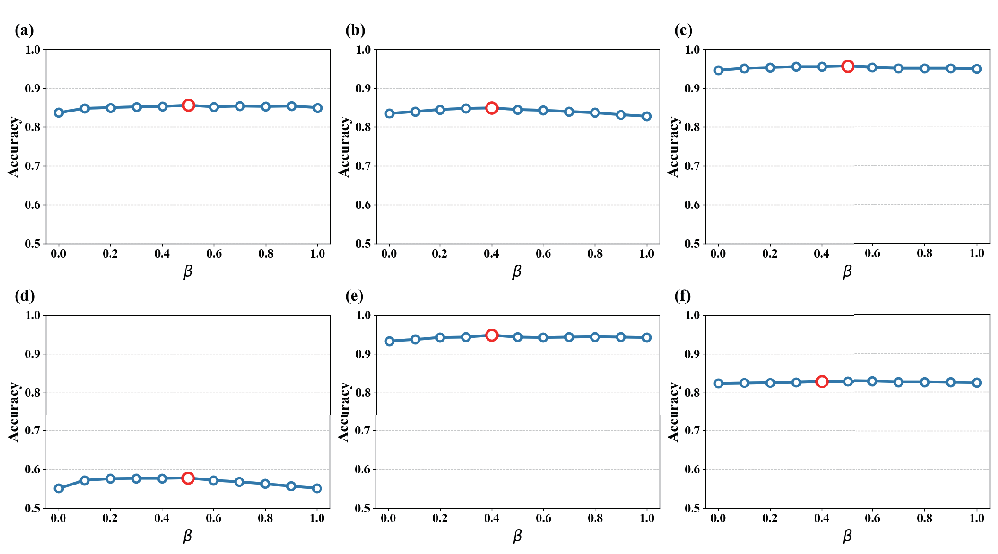}
    \caption{Accuracy of source detection under different values of the parameter $\beta$ and hypergraphs. The red dot indicates the maximum accuracy value within each network. Results are shown for six empirical hypergraphs: (a) Algebra, (b) Restaurants-Rev, (c) Geometry, (d) Email-Eu, (e) Music-Rev, and (f) Bars-Rev.}
    \label{fig:parameter2}
\end{figure}

To examine how the parameter $\beta$ affects the performance of SSDH, we perform a sensitivity analysis under the setting of a 10\% sensor ratio, a 10\% final infection scale, and an infection probability of 0.5. Specifically, for each hypergraph, we measure the detection accuracy as $\beta$ varies from $0$ to $1$. As illustrated in Figs.~\ref{fig:parameter1} and~\ref{fig:parameter2}, the algorithm generally attains its best performance when $\beta=0.5$. We therefore use $\beta=0.5$ as the default parameter choice in all subsequent experiments.

\clearpage
\newpage
\bibliography{apssamp}

@PREAMBLE{
 "\providecommand{\noopsort}[1]{}" 
 # "\providecommand{\singleletter}[1]{#1}%" 
}

@article{williams2005impact,
  title={The impact of HIV/AIDS on the control of tuberculosis in India},
  author={Williams, BG and Granich, R and Chauhan, LS and Dharmshaktu, NS and Dye, C},
  journal={PNAS},
  volume={102},
  number={27},
  pages={9619--9624},
  year={2005},
  publisher={National Academy of Sciences},
  doi = {10.1073/pnas.0501615102}
}

@article{smith2006responding,
  title={Responding to global infectious disease outbreaks: lessons from SARS on the role of risk perception, communication and management},
  author={Smith, Richard D},
  journal={Soc. Sci. Med.},
  volume={63},
  number={12},
  pages={3113--3123},
  year={2006},
  publisher={Elsevier},
  doi = {10.1016/j.socscimed.2006.08.004}
}

@article{antulov2015identification,
  title={Identification of patient zero in static and temporal networks: Robustness and limitations},
  author={Antulov-Fantulin, Nino and Lan{\v{c}}i{\'c}, Alen and {\v{S}}muc, Tomislav and {\v{S}}tefan{\v{c}}i{\'c}, Hrvoje and {\v{S}}iki{\'c}, Mile},
  journal={Phys. Rev. Lett.},
  volume={114},
  number={24},
  pages={248701},
  year={2015},
  publisher={APS},
  doi = {10.1103/PhysRevLett.114.248701}
}

@inproceedings{ru2023inferring,
  title={Inferring patient zero on temporal networks via graph neural networks},
  author={Ru, Xiaolei and Moore, Jack Murdoch and Zhang, Xin-Ya and Zeng, Yeting and Yan, Gang},
  booktitle={Proceedings of the AAAI Conference on Artificial Intelligence},


  pages={9632--9640},
  year={2023},
  publisher={AAAI Press},
  address={Washington, DC},
  doi = {10.1609/aaai.v37i8.26152}
}

@article{fanelli2020analysis,
  title={Analysis and forecast of COVID-19 spreading in China, Italy and France},
  author={Fanelli, Duccio and Piazza, Francesco},
  journal={Chaos, Solitons Fractals},
  volume={134},
  pages={109761},
  year={2020},
  publisher={Elsevier},
  doi = {10.1016/j.chaos.2020.109761}
}

@book{mckay2017patient,
  title={Patient Zero and the Making of the AIDS Epidemic},
  author={McKay, Richard A},
  year={2017},
  publisher={University of Chicago Press},
  doi = {10.7208/chicago/9780226064000.001.0001},
  address = {Chicago}
}

@article{jiang2016identifying,
  title={Identifying propagation sources in networks: State-of-the-art and comparative studies},
  author={Jiang, Jiaojiao and Wen, Sheng and Yu, Shui and Xiang, Yang and Zhou, Wanlei},
  journal={IEEE Commun. Surv. Tutorials},
  volume={19},
  number={1},
  pages={465--481},
  year={2016},
  publisher={IEEE},
  doi = {10.1109/COMST.2016.2615098}
}

@article{lokhov2014inferring,
  title={Inferring the origin of an epidemic with a dynamic message-passing algorithm},
  author={Lokhov, Andrey Y and M{\'e}zard, Marc and Ohta, Hiroki and Zdeborov{\'a}, Lenka},
  journal={Phys. Rev. E},
  volume={90},
  number={1},
  pages={012801},
  year={2014},
  publisher={APS},
  doi = {10.1103/PhysRevE.90.012801}
}

@article{chang2018maximum,
  title={Maximum a posteriori estimation for information source detection},
  author={Chang, Biao and Chen, Enhong and Zhu, Feida and Liu, Qi and Xu, Tong and Wang, Zhefeng},
  journal={IEEE Trans. Syst. Man Cybern.: Syst.},
  volume={50},
  number={6},
  pages={2242--2256},
  year={2018},
  publisher={IEEE},
  doi = {10.1109/TSMC.2018.2811410}
}

@article{li2019locating,
  title={Locating multiple sources of contagion in complex networks under the SIR model},
  author={Li, Xiang and Liu, Yangyang and Zhao, Chengli and Zhang, Xue and Yi, Dongyun},
  journal={Appl. Sci.},
  volume={9},
  number={20},
  pages={4472},
  year={2019},
  publisher={MDPI},
  doi = {10.3390/app9204472}
}

@inproceedings{luo2013estimating,
  title={Estimating infection sources in a network with incomplete observations},
  author={Luo, Wuqiong and Tay, Wee Peng},
  booktitle={2013 IEEE Global Conference on Signal and Information Processing},
  pages={301--304},
  year={2013},
  publisher={IEEE},
  address={Austin, TX},
  doi = {10.1109/GlobalSIP.2013.6736875}
}

@article{zhu2014information,
  title={Information source detection in the SIR model: A sample-path-based approach},
  author={Zhu, Kai and Ying, Lei},
  journal={IEEE/ACM Trans. Networking},
  volume={24},
  number={1},
  pages={408--421},
  year={2014},
  publisher={IEEE},
  doi = {10.1109/TNET.2014.2364972}
}

@article{paluch2020optimizing,
  title={Optimizing sensors placement in complex networks for localization of hidden signal source: A review},
  author={Paluch, Robert and Gajewski, {\L}ukasz G and Ho{\l}yst, Janusz A and Szymanski, Boleslaw K},
  journal={Future Gener. Comput. Syst.},
  volume={112},
  pages={1070--1092},
  year={2020},
  publisher={Elsevier},
  doi = {10.1016/j.future.2020.06.023}
}

@article{pinto2012locating,
  title={Locating the source of diffusion in large-scale networks},
  author={Pinto, Pedro C and Thiran, Patrick and Vetterli, Martin},
  journal={Phys. Rev. Lett.},
  volume={109},
  number={6},
  pages={068702},
  year={2012},
  publisher={APS},
  doi = {10.1103/PhysRevLett.109.068702}
}

@incollection{paluch2020enhancing,
  title={Enhancing maximum likelihood estimation of infection source localization},
  author={Paluch, Robert and Gajewski, {\L}ukasz and Suchecki, Krzysztof and Szyma{\'n}ski, Boles{\l}aw and Ho{\l}yst, Janusz A},
  booktitle={Simplicity of Complexity in Economic and Social Systems: Proceedings of the 54th Winter School of Theoretical Physics},
  pages={21--41},
  year={2020},
  publisher={Springer},
  address={L{\k{a}}dek Zdr{\'o}j, Poland},
  doi = {10.1007/978-3-030-56160-4_2}
}

@article{tang2018estimating,
  title={Estimating infection sources in networks using partial timestamps},
  author={Tang, Wenchang and Ji, Feng and Tay, Wee Peng},
  journal={IEEE Trans. Inf. Forensics Secur.},
  volume={13},
  number={12},
  pages={3035--3049},
  year={2018},
  publisher={IEEE},
  doi = {10.1109/TIFS.2018.2837655}
}

@article{wang2020locating,
  title={Locating source of heterogeneous propagation model by universal algorithm},
  author={Wang, Hong-Jue and Sun, Kai-Jia},
  journal={Europhys. Lett.},
  volume={131},
  number={4},
  pages={48001},
  year={2020},
  publisher={IOP Publishing},
  doi = {10.1209/0295-5075/131/48001}
}

@article{zhu2022locating,
  title={Locating multi-sources in social networks with a low infection rate},
  author={Zhu, Peican and Cheng, Le and Gao, Chao and Wang, Zhen and Li, Xuelong},
  journal={IEEE Trans. Network Sci. Eng.},
  volume={9},
  number={3},
  pages={1853--1865},
  year={2022},
  publisher={IEEE},
  doi = {10.1109/TNSE.2022.3153968}
}

@article{paluch2018fast,
  title={Fast and accurate detection of spread source in large complex networks},
  author={Paluch, Robert and Lu, Xiaoyan and Suchecki, Krzysztof and Szyma{\'n}ski, Boles{\l}aw K and Ho{\l}yst, Janusz A},
  journal={Sci. Rep.},
  volume={8},
  number={1},
  pages={2508},
  year={2018},
  publisher={Nature Publishing Group UK London},
  doi = {10.1038/s41598-018-20546-3}
}

@article{yang2020locating,
  title={Locating the propagation source in complex networks with a direction-induced search based Gaussian estimator},
  author={Yang, Fan and Yang, Shuhong and Peng, Yong and Yao, Yabing and Wang, Zhiwen and Li, Houjun and Liu, Jingxian and Zhang, Ruisheng and Li, Chungui},
  journal={Knowledge-Based Syst.},
  volume={195},
  pages={105674},
  year={2020},
  publisher={Elsevier},
  doi = {10.1016/j.knosys.2020.105674}
}

@inproceedings{wang2022rapid,
  title={A rapid source localization method in the early stage of large-scale network propagation},
  author={Wang, Zhen and Hou, Dongpeng and Gao, Chao and Huang, Jiajin and Xuan, Qi},
  booktitle={Proceedings of the ACM web conference 2022},
  pages={1372--1380},
  year={2022},
  publisher={ACM},
  address={Lyon},
  doi = {10.1145/3485447.3512184}
}

@inproceedings{wang2023lightweight,
  title={Lightweight source localization for large-scale social networks},
  author={Wang, Zhen and Hou, Dongpeng and Gao, Chao and Li, Xiaoyu and Li, Xuelong},
  booktitle={Proceedings of the ACM web conference 2023},
  pages={286--294},
  year={2023},
  publisher={ACM},
  address={Austin, TX},
  doi = {10.1145/3543507.3583299}
}

@article{cheng2024sdsi,
  title={A heuristic framework for sources detection in social networks via graph convolutional networks},
  author={Cheng, Le and Zhu, Peican and Gao, Chao and Wang, Zhen and Li, Xuelong},
  journal={IEEE Trans. Syst. Man Cybern.: Syst.},
  year={2024},
  volume={54},
  pages={7002--7014},
  publisher={IEEE},
  doi = {10.1109/TSMC.2024.3448226}
}

@article{ali2020revisit,
  title={A revisit to the infection source identification problem under classical graph centrality measures},
  author={Ali, Syed Shafat and Anwar, Tarique and Rizvi, Syed Afzal Murtaza},
  journal={Onl. Soc. Netw. Media},
  volume={17},
  pages={100061},
  year={2020},
  publisher={Elsevier},
  doi = {10.1016/j.osnem.2020.100061}
}

@article{hu2023source,
  title={Source localization in complex networks with optimal observers based on maximum entropy sampling},
  author={Hu, Zhao-Long and Wang, Hong-Jue and Sun, Lei and Tang, Chang-Bing and Li, Minglu},
  journal={Expert Syst. Appl.},
  volume={256},
  pages={124946},
  year={2024},
  publisher={Elsevier},
  doi = {10.2139/ssrn.4655452}
}

@article{cheng2025efficient,
  title={Efficient source detection in incomplete networks via sensor deployment and source approaching},
  author={Cheng, Le and Zhu, Peican and Tang, Keke and Gao, Chao and Wang, Zhen},
  journal={IEEE Trans. Inf. Forensics Secur.},
  year={2025},
  volume={20},
  pages={3705--3716},
  publisher={IEEE},
  doi = {10.1109/TIFS.2025.3550069}
}

@article{kim2024higher,
  title={Higher-order components dictate higher-order contagion dynamics in hypergraphs},
  author={Kim, Jung-Ho and Goh, K-I},
  journal={Phys. Rev. Lett.},
  volume={132},
  number={8},
  pages={087401},
  year={2024},
  publisher={APS},
  doi = {10.1103/PhysRevLett.132.087401}
}

@article{bick2023higher,
  title={What are higher-order networks?},
  author={Bick, Christian and Gross, Elizabeth and Harrington, Heather A and Schaub, Michael T},
  journal={SIAM Rev.},
  volume={65},
  number={3},
  pages={686--731},
  year={2023},
  publisher={SIAM},
  doi = {10.1137/21M1414024}
}

@article{majhi2022dynamics,
  title={Dynamics on higher-order networks: A review},
  author={Majhi, Soumen and Perc, Matja{\v{z}} and Ghosh, Dibakar},
  journal={J. R. Soc. Interface},
  volume={19},
  number={188},
  pages={20220043},
  year={2022},
  publisher={The Royal Society},
  doi = {10.1098/rsif.2022.0043}
}

@misc{ke2025source,
  title={Source Detection in Hypergraph Epidemic Dynamics using a Higher-Order Dynamic Message Passing Algorithm},
  author={Ke, Qiao and Masuda, Naoki and Jin, Zhen and Liu, Chuang and Zhan, Xiu-Xiu},
  eprint={2507.02523},
  archivePrefix={arXiv},
  year={2025}
}

@misc{cheng2025hyperdet,
  title={HyperDet: Source Detection in Hypergraphs via Interactive Relationship Construction and Feature-rich Attention Fusion},
  author={Cheng, Le and Zhu, Peican and Guo, Yangming and Tang, Keke and Gao, Chao and Wang, Zhen},
  eprint={2505.12894},
  archivePrefix={arXiv},
  year={2025}
}

@article{yu2024source,
  title={Source inference for misinformation spreading on hypergraphs},
  author={Yu, Xiaohang and Nie, Yanyi and Li, Wenyao and Luo, Ganzhi and Lin, Tao and Wang, Wei},
  journal={Chaos, Solitons Fractals},
  volume={187},
  pages={115457},
  year={2024},
  publisher={Elsevier},
  doi = {10.1016/j.chaos.2024.115457}
}

@article{surana2022hypergraph,
  title={Hypergraph similarity measures},
  author={Surana, Amit and Chen, Can and Rajapakse, Indika},
  journal={IEEE Trans. Network Sci. Eng.},
  volume={10},
  number={2},
  pages={658--674},
  year={2022},
  publisher={IEEE},
  doi = {10.1109/TNSE.2022.3217185}
}

@article{feng2024hyper,
  title={A hyper-distance-based method for hypernetwork comparison},
  author={Feng, Ruonan and Xu, Tao and Xie, Xiaowen and Zhang, Zi-Ke and Liu, Chuang and Zhan, Xiu-Xiu},
  journal={Chaos},
  volume={34},
  pages = {083120},
  number={8},
  year={2024},
  publisher={AIP Publishing},
  doi = {10.1063/5.0221267}
}

@article{watts1998collective,
  title={Collective dynamics of ‘small-world’networks},
  author={Watts, Duncan J and Strogatz, Steven H},
  journal={Nature},
  volume={393},
  number={6684},
  pages={440--442},
  year={1998},
  publisher={Nature Publishing Group},
  doi = {10.1038/30918}
}

@article{xie2023efficient,
  title={An efficient adaptive degree-based heuristic algorithm for influence maximization in hypergraphs},
  author={Xie, Ming and Zhan, Xiu-Xiu and Liu, Chuang and Zhang, Zi-Ke},
  journal={Inf. Process. Manage.},
  volume={60},
  number={2},
  pages={103161},
  year={2023},
  publisher={Elsevier},
  doi = {10.1016/j.ipm.2022.103161}
}

@misc{amburg2020fair,
  title={Hypergraph clustering for finding diverse and experienced groups},
  author={Amburg, Ilya and Veldt, Nate and Benson, Austin R},
  eprint={2006.05645},
  archivePrefix={arXiv},
  year={2020}
}

@inproceedings{ni2019justifying,
  title={Justifying recommendations using distantly-labeled reviews and fine-grained aspects},
  author={Ni, Jianmo and Li, Jiacheng and McAuley, Julian},
  booktitle={Proceedings of the 2019 conference on empirical methods in natural language processing and the 9th international joint conference on natural language processing (EMNLP-IJCNLP)},
  pages={188--197},
  year={2019},
  publisher={ACL},
  address={Hong Kong},
  doi = {10.18653/v1/D19-1018}
}

@inproceedings{yin2017local,
  title={Local higher-order graph clustering},
  author={Yin, Hao and Benson, Austin R and Leskovec, Jure and Gleich, David F},
  booktitle={Proceedings of the 23rd ACM SIGKDD international conference on knowledge discovery and data mining},
  pages={555--564},
  year={2017},
  publisher={ACM},
  address={Halifax, NS},
  doi = {10.1145/3097983.3098069}
}

@article{feng2025hypergraph,
  title={Hypergraph dismantling with spectral clustering},
  author={Feng, Ruonan and Ke, Qiao and She, Li and Kong, Xiangjie and Liu, Chuang and Zhan, Xiu-Xiu},
  journal={Commun. Nonlinear Sci. Numer. Simul.},
  pages={108975},
  volume={150},
  year={2025},
  publisher={Elsevier},
  doi = {10.1016/j.cnsns.2025.108975}
}

@article{li2011attack,
  title={Attack robustness of scale-free networks based on grey information},
  author={Li, Jun and Wu, Jun and Li, Yong and Deng, Hong-Zhong and Tan, Yue-Jin},
  journal={Chin. Phys. Lett.},
  volume={28},
  number={5},
  pages={058904},
  year={2011},
  publisher={IOP Publishing},
  doi = {10.1088/0256-307X/28/5/058904}
}

@article{yao2023lesson,
  title={Lesson learned from COVID-19 retrospective study: An entropy-based clinical-interpretable scorecard for mortality risk control at ICU admission},
  author={Yao, Chong and Huangqi, Chonghui and Huang, Anpeng},
  journal={Tsinghua Sci. Technol.},
  volume={29},
  number={1},
  pages={34--45},
  year={2024},
  publisher={TUP},
  doi = {10.26599/TST.2023.9010042}
}

@article{xu2023iqabc,
  title={Iqabc-based hybrid deployment algorithm for mobile robotic agents providing network coverage},
  author={Xu, Shuang and Liu, Xiaojie and Li, Dengao and Zhao, Jumin},
  journal={Tsinghua Sci. Technol.},
  volume={29},
  number={2},
  pages={589--604},
  year={2024},
  publisher={TUP},
  doi = {10.26599/TST.2023.9010074}
}

@article{brockmann2013hidden,
  title={The hidden geometry of complex, network-driven contagion phenomena},
  author={Brockmann, Dirk and Helbing, Dirk},
  journal={Science},
  volume={342},
  number={6164},
  pages={1337--1342},
  year={2013},
  publisher={American Association for the Advancement of Science},
  doi = {10.1126/science.1245200}
}

@article{rombach2014core,
  title={Core-periphery structure in networks},
  author={Rombach, M Puck and Porter, Mason A and Fowler, James H and Mucha, Peter J},
  journal={SIAM J. Appl. Math.},
  volume={74},
  number={1},
  pages={167--190},
  year={2014},
  publisher={SIAM},
  doi = {10.1137/120881683}
}

@article{Ali2025acc,
  title={Generalized Local Prominence for Source Detection in Real-World Rumor Networks},
  author={Ali, Syed Shafat and Rastogi, Ajay and Anwar, Tarique and Rizvi, Syed Afzal M and Yang, Jian and Wu, Jia and Sheng, Quan Z},
  journal={IEEE Trans. Knowl. Data Eng.},
  year={2025},
  volume={37},
  pages = {4620--4634},
  publisher={IEEE},
  doi = {10.1109/TKDE.2025.3567282}
}

@article{zhao2024aed,
  title={MASE: Multi-attribute source estimator for epidemic transmission in complex networks},
  author={Zhao, Jie and Cheong, Kang Hao},
  journal={IEEE Trans. Syst. Man Cybern.: Syst.},
  volume={54},
  number={6},
  pages={3308--3320},
  year={2024},
  publisher={IEEE},
  doi = {10.1109/TSMC.2024.3349537}
}

@article{Geon2023,
  title={Temporal hypergraph motifs},
  author={Lee, Geon and Shin, Kijung},
  journal={Knowl. Inf. Syst.},
  volume={65},
  number={4},
  pages={1549--1586},
  year={2023},
  publisher={Springer},
  doi = {10.1007/s10115-023-01837-2}
}

@article{Neuh2021,
  title={Consensus dynamics on temporal hypergraphs},
  author={Neuh{\"a}user, Leonie and Lambiotte, Renaud and Schaub, Michael T},
  journal={Phys. Rev. E},
  volume={104},
  number={6},
  pages={064305},
  year={2021},
  publisher={APS},
  doi = {10.1103/PhysRevE.104.064305}
}

@article{pan2024robustness,
  title={Robustness of interdependent hypergraphs: A bipartite network framework},
  author={Pan, Xingyu and Zhou, Jie and Zhou, Yinzuo and Boccaletti, Stefano and Bonamassa, Ivan},
  journal={Phys. Rev. Res.},
  volume={6},
  number={1},
  pages={013049},
  year={2024},
  publisher={APS},
  doi={10.1103/PhysRevResearch.6.013049}
}

@article{jhun2021effective,
  title={Effective epidemic containment strategy in hypergraphs},
  author={Jhun, Bukyoung},
  journal={Phys. Rev. Res.},
  volume={3},
  number={3},
  pages={033282},
  year={2021},
  publisher={APS},
  doi={10.1103/PhysRevResearch.3.033282
}
}

@article{zhan2025measuring,
  title={Measuring and utilizing temporal network dissimilarity},
  author={Zhan, Xiu-Xiu and Liu, Chuang and Wang, Zhipeng and Wang, Huijuan and Holme, Petter and Zhang, Zi-Ke},
  journal={Commun. Phys.},
  volume={8},
  number={1},
  pages={40},
  year={2025},
  publisher={Nature Publishing Group UK London},
  doi={10.1038/s42005-025-01940-6}
}

@article{zhang2025locating,
  title={Locating influential nodes in hypergraphs via fuzzy collective influence},
  author={Zhang, Su-Su and Yu, Xiaoyan and Sun, Gui-Quan and Liu, Chuang and Zhan, Xiu-Xiu},
  journal={Commun. Nonlinear Sci. Numer. Simul.},
  volume={142},
  pages={108574},
  year={2025},
  publisher={Elsevier},
  doi={10.1016/j.cnsns.2024.108574}
}

\end{document}